\documentclass[preprint,amsmath,amssymb,aps,prb,superscriptaddress,showpacs,citeautoscript,floatfix]{revtex4-1}
\usepackage[pdftex]{graphicx}
\usepackage{dcolumn}
\usepackage{bm}
\usepackage{amsmath}
\usepackage{amssymb}
\usepackage[]{units}
\usepackage{booktabs}
\usepackage{epstopdf}
\usepackage{color,soul}
\usepackage{subfigure}

\begin{document}

\title{The electronic and magnetic structure of p-element (C,N) doped rutile-TiO$_{2}$; A hybrid DFT study}

\author{J Atanelov}
\email{ja@cms.tuwien.ac.at}
\affiliation{Institute of Applied Physics, Vienna University of Technology, Gu\ss hausstra\ss e 25-25a, 1040 Vienna, Austria}
\author{C Gruber}
\affiliation{Institute of Applied Physics, Vienna University of Technology, Gu\ss hausstra\ss e 25-25a, 1040 Vienna, Austria}
\author{P Mohn}
\affiliation{Institute of Applied Physics, Vienna University of Technology, Gu\ss hausstra\ss e 25-25a, 1040 Vienna, Austria}

\begin{abstract}

We study the electronic and magnetic structure of carbon and nitrogen impurities and interstitials in rutile TiO${_2}$. To this end we perform \textit{ab-initio} calculations of a 48-atom supercell employing the VASP code. In order to obtain a realistic description of the electronic and magnetic structure, exchange and correlation are treated with the HSE06 hybrid functional. Both, atomic positions and cell dimensions are fully relaxed. Substitutional carbon and nitrogen are found to have a magnetic moment of 2 and 1$\mu{_B}$, respectively, with a tendency for anti-ferromagnetic long range order. For C/N on interstitial sites we find that carbon is non-magnetic while nitrogen always possesses a magnetic moment of 1$\mu{_B}$. We find that these interstitial positions are on a saddle point of the total energy. The stable configuration is reached when both carbon and nitrogen form a C-O and N-O dimer with a bond length close to the double bond for CO and NO. This result is in agreement with earlier experimental investigations detecting such N-O entities from XPS measurements. The frequencies of the symmetric stretching mode are calculated for these dimers, which could provide a means for experimental verification. For all configurations investigated both C and N states are found inside the TiO${_2}$ gap. These new electronic states are discussed with respect to tuning doped TiO${_2}$ for the application in photocatalysis.     
\end{abstract}


\maketitle

\section{Introduction}

TiO$_{2}$ plays an important role in the fields of dilute magnetic semiconductors (DMS) and photocatalysis. For both applications TiO$_{2}$ requires tailor made electronic properties, which can be manipulated by selectively diluting the system by insertion of dopants.

DMS have been investigated in the past in order to gain new insights in the functional principles and the manipulation of the magnetic and semiconducting properties. The search for magnetic semiconductors or for half-metals used for spin-injection led to the discovery of the new material class of $p$-electron magnets. The prospect of controlling the charge and the electron spin as information carriers would make it possible to combine information processing and storage at the same time \cite{Ohno1998,MatsukuraOhno1998,DiVincenzo1998}. The second major application is in the field of photocatalysis, where the favourable electronic gap range of $1.65-3.1$ eV can be also achieved by doping \cite{VarleyJanottiWalle2011,WangLiu2012,GoshEnglish2012} . 

Regarding rutile TiO$_{2}$ from a photocatalytic point of view we are especially interested in the manipulation of the band gap by incorporation of impurities. 
Rutile TiO$_{2}$ has a band gap of $3.0$ eV, which corresponds to an absorption spectrum in the ultraviolet region. The absorption of photons lead to the creation of electron hole pairs, which are of special relevance for photocatalytic and photoelectrochemical applications like the photo-induced decomposition of water on
TiO$_{2}$ electrodes demonstrated by Fujishima and Honda \cite{Fujishima1972}. Another line of research regards the photocatalytic properties; due do the creation of electron hole pairs the valence band (VB) becomes oxidative and the conduction band (CB) reductive. Surrounding molecules can therefore be oxidized or reduced and as a consequence form radicals which are harmful to organic compounds like bacteria and fungi. Hence, TiO$_{2}$ can be used as a purifier, removing gaseous or aqueous contaminants \cite{Fox1993,Hoffmann2006}. Beside water- and air-purification there is a wide range of other photocatalytic applications, like antifogging and self-cleaning surfaces \cite{Fujishima1998}. Since UV light only makes up 5\% of the sunlight spectrum one aims to shift the absorption spectrum of TiO$_{2}$ into the visible region to improve the absorption rate and therefore the photocatalytic efficiency. Several studies have shown that impurities like nitrogen and carbon verifiable modify the band gap. Beside computational studies done by DiValentin et al. \cite{DiValentin2004a,DiValentin2005} and Yang et al. \cite{Yang2006} there are experimental studies confirming the earlier theoretical results. Diwald et al. \cite{Diwald2004a} reported a blueshift in the band gap after doping with nitrogen. In a second study \cite{Diwald2004b} they observed a redshift by inducing nitrogen into the TiO$_{2}$ rutile host matrix. For the first case they assumed oxygen atoms being substituted by nitrogen impurities and for the latter nitrogen atoms being placed on interstitial sites. Batzill et al. \cite{Batzill2006} on the other hand reported to see a redshift when substituting O by N. Motivated by these partly contradictory reports we perform our theoretical investigation in order to interpret the experimental results.

In the present study we investigate TiO$_{2}$ rutile doped with carbon and nitrogen atoms, respectively, where we assume C and N either to replace oxygen or to occupy interstitial positions. Although magnetic order is most common in metallic materials with narrow bands of $d$- or $f$- electrons, the carriers of the magnetic moments in doped semiconductors or insulators like TiO$_{2}$ are the carbon or nitrogen atoms. The magnetic moment is produced by the $p$-electrons which become polarized because of the flat $p$-bands of these impurity atoms \cite{Dietl2000}. This is called $p$-electron magnetism and has been investigated intensely during the last years.

\section{Crystal Structure and Doping Configurations}

There are three common polymorphs of Titanium dioxide: rutile,
brookite and anatase. Rutile is the thermodynamically most stable modification of TiO$_{2}$ and is an indirect wide band gap semiconductor with an experimental band gap of $3.0$ eV \cite{Grant1959}. 

The rutile structure belongs to the P4$_{2}$/mnm (No. 136) tetragonal space group with unit cell parameters of $a=b=4.587$ \AA\ and $c=2.954$ \AA.
In rutile TiO$_{2}$ every titanium atom is octahedrally coordinated to six oxygen atoms. The so formed octahedrons show an orthorhombic distortion, with the apical Ti-O bond
length being slightly longer than the equatorial Ti-O bond length. Each TiO$_{6}$ octahedron is in contact with 10 neighbour octahedrons. The TiO$_{2}$ rutile crystal structure therefore can be seen as a chain of edge and corner-sharing TiO$_{6}$ units \cite{Grant1959,Graetzel1996}.

\begin{figure}[h]
   \centering
   \subfigure{\includegraphics[width=0.45\linewidth]{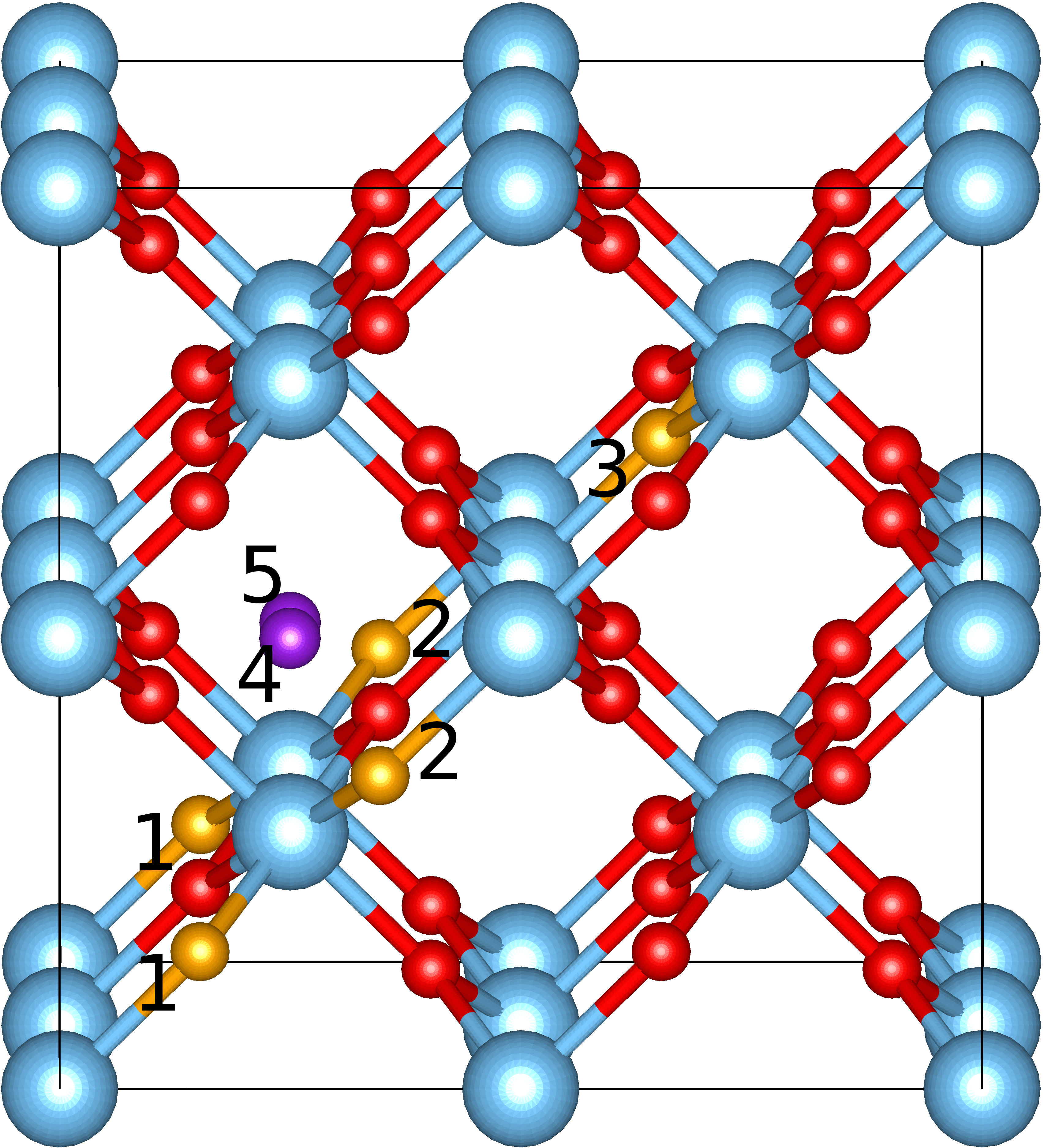}} \quad
   \subfigure{\includegraphics[width=0.50\linewidth]{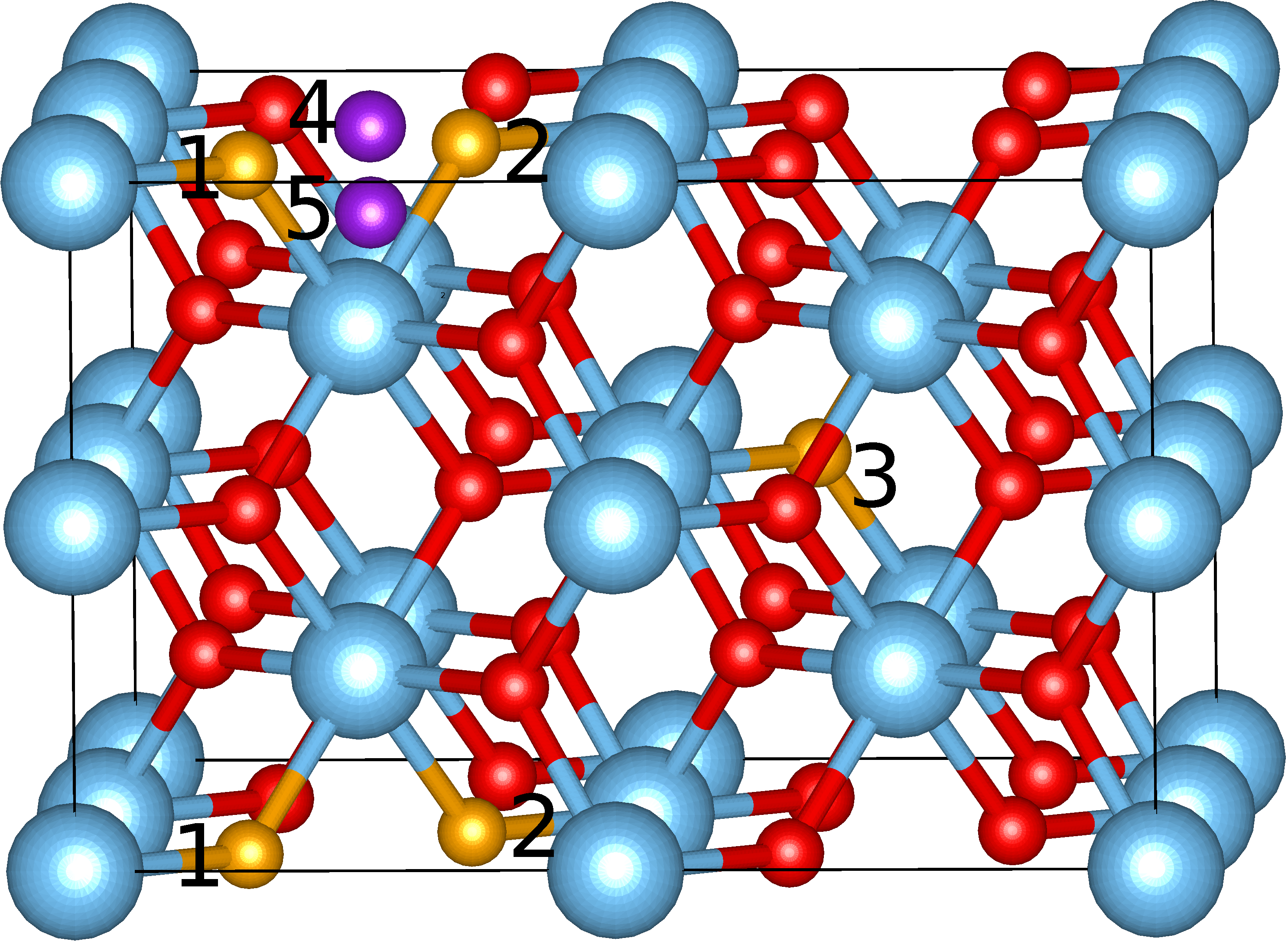}}\\
  \caption{(a) and (b) shows the 48-atom supercell of TiO$_{2}$. The large blue and the small red spheres represent the Ti and the O atoms. The orange numbered spheres (1-3) denote the positions of O substituted by the C and N dopants and the violet numbered spheres (4-5) notify the interstitial doping positions.}
  \label{fig:supercell}
\end{figure}

Beside the investigation of the electronic properties of pristine TiO$_{2}$, we examined the effect of C and N atoms incorporated in various concentrations and positions into the host matrix.  In total seven different levels and configurations of doping were examined using \textit{ab-initio} calculations.
Fig.\ref{fig:supercell} depicts the different positions of the C and N dopants in the TiO$_{2}$ $2\times2\times2$ (48 atoms) supercell. In one of the performed calculations we assumed a single oxygen atom of the host matrix to be replaced by C or N (position (1)). This corresponds to a doping concentration of $2.1{\%}$. Further, two substitutional oxygen sites with two C or two N atoms, respectively, were occupied leading to a doping rate of $4.2{\%}$. This was done for two different configurations investigating the effect of increasing distance between the two impurity atoms. In the first configuration of this double substitutional state the distance between the two dopants was chosen to be minimal (position (1) and (2)). In the second one the two dopants were placed on arbitrary oxygen sites (positions (1) and (3)) increasing the distance. Moreover, two different interstitial positions of C and N (positions (4) and (5)) in the TiO$_{2}$ host lattice were examined. Finally, we performed a set of calculations exchanging these interstitial dopants by their next nearest oxygen neighbour, so that O and C or N respectively switch position, with a doping concentration of again $2.1{\%}$.

\section{Calculational Details}

All calculations were performed employing the \textit{Vienna Ab-initio Simulation Package} (VASP) \cite{Kresse1993, Kresse1994, Kresse1994CondMa, Kresse1996Comp, Kresse1996PhysRev, KresseJoubert1999} which uses projector augmented wave (PAW) pseudopotentials \cite{Bloechl1994} to describe the potential between the ions. The semi-core $s$ and $p$ states for Ti and the $2s$ states for O, C and N respectively were incorporated in the calculations. To save computation time, the effects of exchange and correlation were initially treated using the GGA-PBE approximation \cite{PBE1996,PBE1997}, 
until all force components were smaller than $0.01$ eV/\AA. During the relaxation we allowed for a change of the atomic positions, shape and volume of the cell. 
The final convergence was done with the post-DFT HSE06 functional \cite{Heyd2006} which is known to yield reliable results for the gap size and the position of the impurity bands inside the gap. 

Plane waves with an energy up to $530$ eV were included in the basis set, in order to avoid Pulay stress and other related problems. The Brillouin-Zone integration was performed using a $4\times4\times4$ $\Gamma$ centered \textit{k}-mesh with Gaussian smearing set to $0.05$ eV. The total energy was converged better than $1\times10^{-6}$ eV for all cases investigated. 
The limitations of LDA and GGA+U in predicting the equilibrium lattice constant and the proper band gap  \cite{Arroyo2011}
poses problems in the description of the structural and electronic properties of the pristine and doped system. Thus, predicting the position of the impurity states, crucial for photocatalytic reasons, becomes only possible after extracting the appropriate U from experimental data.

\section{C/N-substitutions}

To compare and evaluate the structural changes induced by the impurities calculations on pure TiO$_{2}$ were done initially to obtain structural and electronic data that can be used as a reference. The calculated lattice parameters 
of pure TiO$_{2}$ rutile (  $a=4.65\AA$ ; $c=2.97\AA$ ) are in good agreement with the experimental values ( $a=4.59\AA $ ; $c=2.95\AA$ ). The direct band gaps calculated using the HSE06 functional ( $3.24eV$ ) show a considerably improvement over the GGA results ( $1.77eV$ ) \cite{Kresse2010} and are again in in good agreement with experimental values ( $3.1eV$ )
\cite{Grant1959}.

\subsection{Single Carbon and Nitrogen Substitution}

In total three substitutional cases were investigated. i) one single oxygen becomes substituted by C or N impurity. Further, two oxygen sites are substituted by either two carbons or two nitrogens, considering two different cases ii) minimizing the distance of the impurities occupying two adjacent oxygen sites, and iii) the impurities occupy two distant oxygen positions. The latter two cases allow to study the dependence of the magnetic coupling on the distance of the magnetic ions.

\begin{figure}[h]
  \centering
  \includegraphics[width=0.5\textwidth, angle=-90]{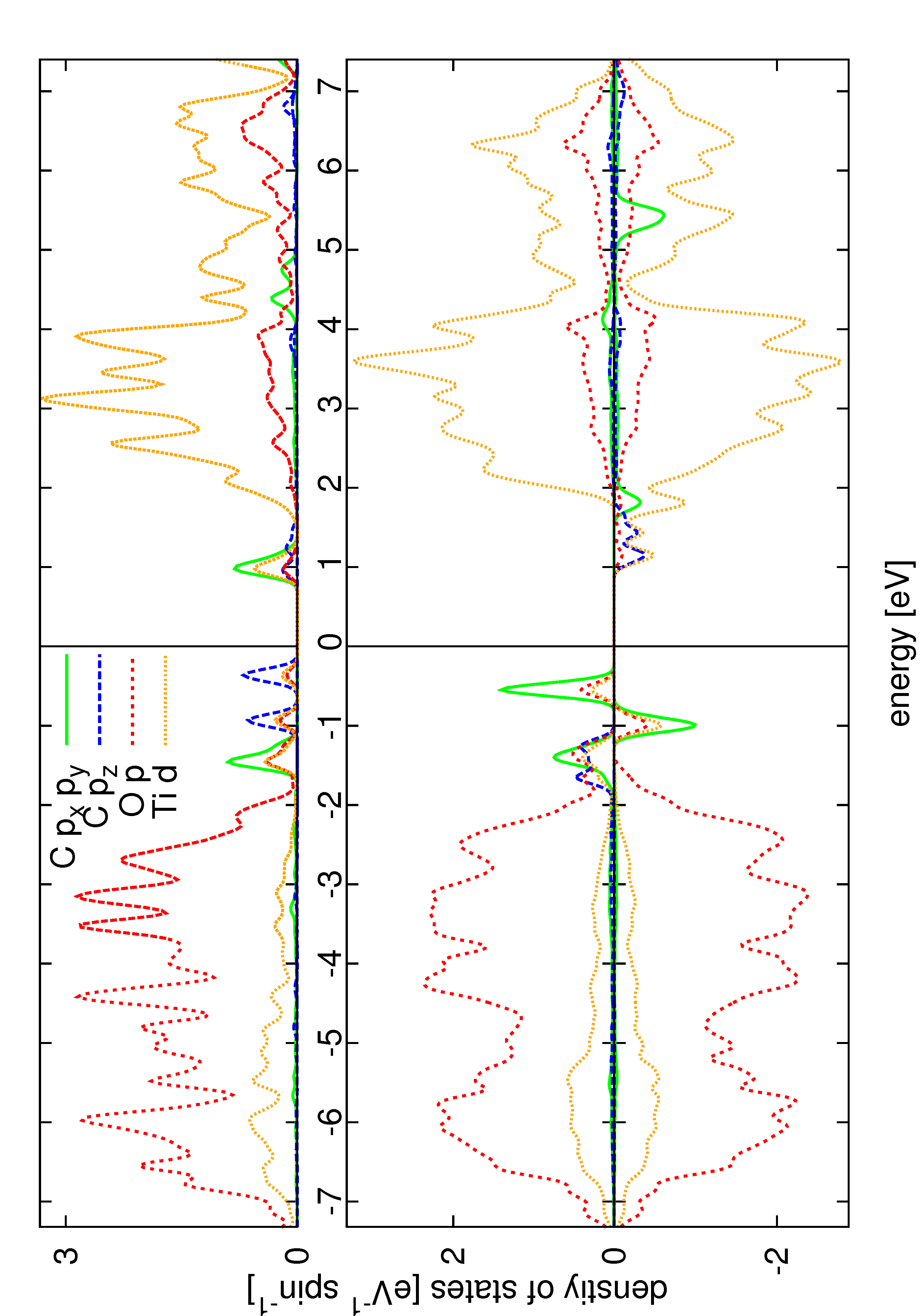}
  \caption{Density of states of the non-magnetic (upper panel) and magnetic (lower panel) state of C substituted TiO$_{2}$. The DOS shown includes only the NN around the C impurity, thus 2 Ti and 5 O atoms.}
  \label{fig:oneC}
\end{figure}

Substituting just one oxygen atom by a carbon impurity leads to a magnetic moment of $2\mu_{B}$ per supercell and arises mainly from the C atom with $0.717\mu_{B}$ per carbon atom (measured within the muffin tin radius 1.63 \AA). Substituting oxygen by carbon introduces two holes into the system. As a consequence one complete band on the average becomes unoccupied leading to a semiconducting ground state but with a magnetic moment. The total moment of $2\mu_{B}$ per unit cell is composed of the magnetic moment at the carbon site plus the polarizations of the surrounding oxygen.       
Figure \ref{fig:oneC} depicts the density of states (DOS) for the single substitutional case. The upper panel shows the non-magnetic case of carbon substitution. The lower one depicts the DOS of the equilibrium magnetic state with $2\mu_{B}$ .

\begin{figure}[h]
  \centering
  \includegraphics[width=0.5\textwidth, angle=-90]{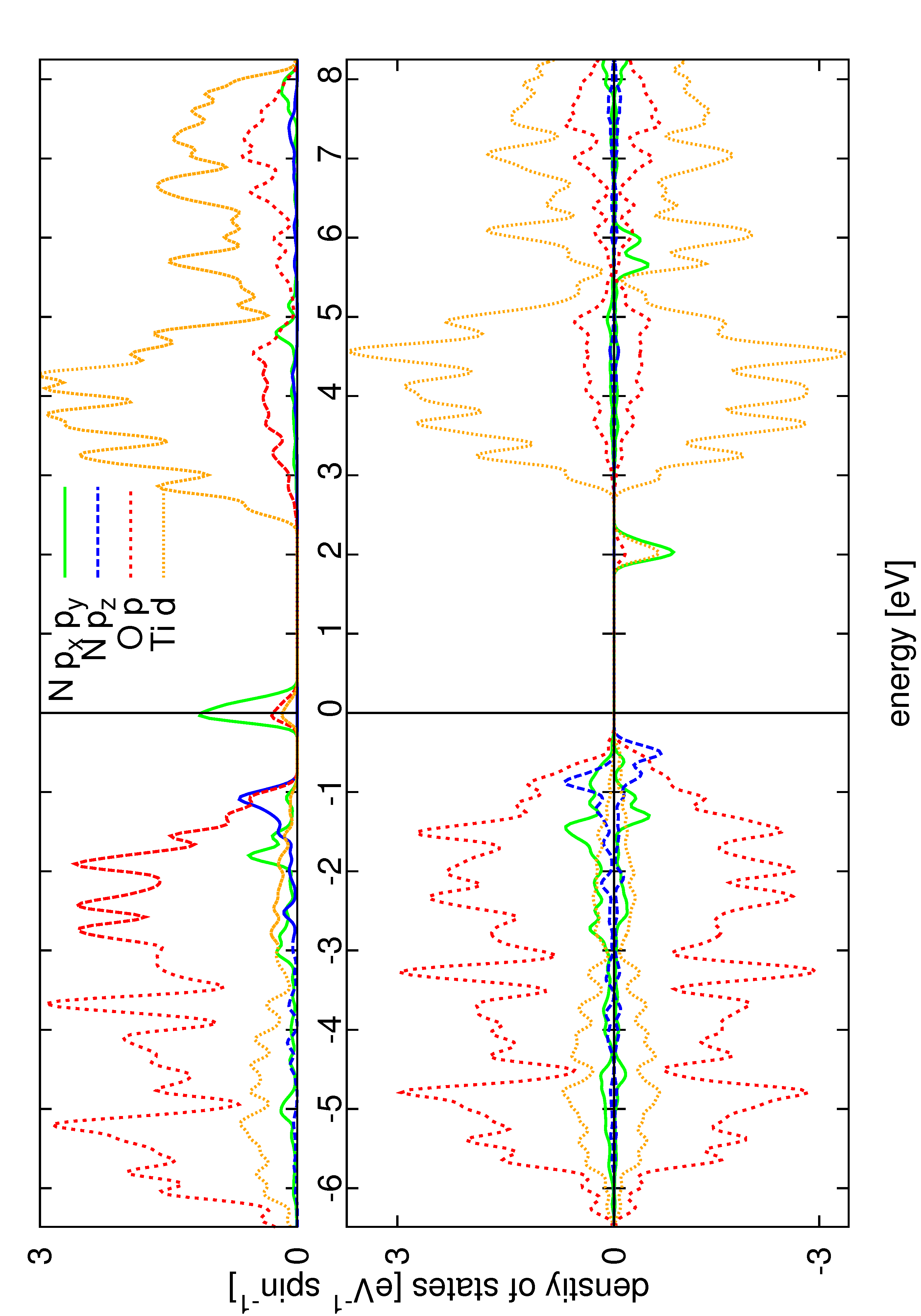}
  \caption{Density of states of the non-magnetic (upper panel) and magnetic (lower panel) state of N substituted TiO$_{2}$. The DOS shown includes only the next neighbours around the N impurity, thus 2 Ti and 5 O atoms.}
  \label{fig:oneN}
\end{figure}

The substitution of N introduces one hole and consequently leads to a magnetic moment of $1\mu_{B}$  per supercell where $0.57\mu_{B}$ are located at the N site (measured within the muffin tin radius 1.40 \AA). This result agrees with the experimental finding of room temperature ferromagnetism in N-doped rutile TiO$_{2}$ films by Bao et al.\cite{Bao2011} who report a nitrogen moment of about $0.9\mu_{B}$. The DOS for the magnetic and the non-magnetic state is shown in Figure \ref{fig:oneN}. 

Introducing dopants into the TiO$_{2}$ rutile crystal leads to a change in the atomic distances. Changes of the bond lengths between the oxygen (of the undoped system), the substituted carbon and the substituted nitrogen atom to its next neighbours (NN), respectively, were compared. The atomic distances of the pristine crystal lattice were normalized to those of the doped systems using its relaxed volume. By doing so we considered increased distances caused by the larger ionic volume of the C and N dopant. Table \ref{table:oneimp} gives an overview on the bond lengths in the respective system. Both the N and the C atom causes an enlargement of the atomic distances in the crystal. Due to the higher electronegativity of N($3.04$)$>$C($2.55$) \cite{Allred1961}, nitrogen tends to interact more with the host semiconductor in particular with the neighbouring oxygens. This leads to shorter bond lengths for N to its next neighbours than for C. These results are again in line with the DOS shown in figure 3 and 4. The larger bond lengths and the consequently weaker bonding to the host semiconductor makes the $2p$ states of the C atom more localized in the gap and the C states which occur in the VB and CB due to the weaker interaction with O are smaller than in the N case. The $2p$ states of nitrogen are mostly located in the VB and CB, induced by the stronger bonding. As a consequence carbon doped TiO$_{2}$ rutile has a smaller band gap than the nitrogen doped system (table \ref{table:oneimp}).

To investigate the stability of the doped system we calculated the defect formation energies according to the following formula \cite{VandeWalle2004}:
\begin{equation}
  E^{form}[X]=E_{tot}[X]-E_{tot}[bulk]+n(\mu_{O}-\mu_{X})
\end{equation}
where $E_{tot}$ is the total energy of the doped supercell with one oxygen atom replaced by the impurity X(X=C,N), $E_{tot}[bulk]$ the total energy for the pristine system. $n$ indicates the number of oxygen atoms that have been replaced in the supercell by dopant atoms, $\mu_{O}$ 
and $\mu_{X}$ are the corresponding chemical potentials. 

The stability of the various configurations differ with the oxygen chemical potential. $\mu_{O}$ describes the oxygen environment during synthesis
and therefore effects the defect formation energies. We define the oxygen chemical potential being $\mu_{O}$=$\frac{1}{2}\mu_{O_{2}}$+$\mu_{O'}$ with $\mu_{O'}$ ranging from 0 eV to -4 eV, whereas $\mu_{O'}$=0 eV defines the oxygen rich and $\mu_{O'}$=-4 eV the oxygen poor case. The oxygen poor case is approximately half the HSE calculated formation enthalpy of rutile TiO$_{2}$ with $\Delta H_{f}(TiO_{2})=-9.95 eV$ which is in good agreement with the experimental value of 9.80 eV \cite{KubaschewskiO}.   
To give a more conceptual measure of the oxygen concentration we convert $\mu_{O'}$ to oxygen pressure at a temperature of 1000K (top x-axis), typical for annealing of rutile TiO$_{2}$ \cite{Schefflerreuter100}.

In the tables below we list the defect formation energy for the different doping configurations considering the oxygen rich case. The chemical potentials were calculated with respect to C in Diamond, N in N$_{2}$ and O in O$_{2}$. In addition  phase diagrams and $E^{form}[X]$/$\mu_{O'}$ plots are presented in Fig.\ref{fig:phasediagram}.

\begin{table}[h]
\centering
\caption{Single carbon and nitrogen substitution: Cell volume, bond lengths of X-Ti$_{equatorial/axial}$ and X-O$_{[2]}$, total magnetic Moment $M_{tot}$ of the supercell, defect formation energy $E^{form}[X]$, energy difference of the non spin polarized and spin polarized state $\Delta E[NSP-SP]$ and band gap for spin-up and spin-down.}
\renewcommand{\arraystretch}{1.15}
\begin{tabular}{lcccc} 
	X=   & C	& N  & O	\\
	\hline
	Cell Volume (\AA$^{3}$)  	& 519.95 (+1.9\%)  & 516.18 (+1.1\%)        & 510.06 \\
	X-Ti$_{equatorial}$ (\AA)   & 2.07 (+2.8\%) & 2.02 (+0.4\%) & 2.00 \\
	X-Ti$_{axial}$ (\AA)   		& 2.13 (+7.9\%) & 2.02 (+2.6\%) & 1.96 \\
	X-O$_{[2]}$ (\AA)   		& 2.74 (+6.0\%) & 2.62 (+1.9\%) & 2.57 \\
	$M_{tot}$ ($\mu_{B}$)   	& 2             & 1             & 0 \\
    $E^{form}[X]$ (eV)          & 9.42          & 5.43          & -- \\	
	$\Delta E[NSP-SP]$ (eV)   	& 0.24          & 0.80          & -- \\  
 	Spin up gap (eV)         	& 2.20		    & 3.20 	        & 3.24 \\
 	Spin down gap (eV)   		& 2.06          & 2.60          & 3.24 \\
\label{table:oneimp}
\end{tabular}
\end{table}

\subsection{Multiple Carbon and Nitrogen substitution}

As a second step we study the substitution of 2 oxygens by 2 carbons or 2 nitrogens at different but crystallographically equivalent sites. The results are given in table \ref{table:twoimp} and the positions of the substituted atoms correspond with the site numbering given in Fig.\ref{fig:supercell}.	For all 4 cases investigated we find a magnetically ordered ground state with 4$\mu_B$ per supercell for 2 carbons and 2$\mu_B$ per supercell for 2 nitrogens. The additional substitution leads to a further increase of the cell volume, but considerably smaller than for the single impurity (compare to table \ref{table:oneimp}). The formation energy per atom remains almost constant and in general shows a slight increase with respect to the single impurity, only in the case of 2 distant nitrogens a very small reduction is found. Studying 2 impurities gives us the opportunity to discriminate between a ferromagnetic (FM) and an anti-ferromagnetic (AFM) coupling. In general we find the AFM state to be lower in energy, only for the case of 2 distant nitrogens an extremely small FM stabilization energy has been calculated, however, this small energy change is at the verge of the numerical accuracy. As expected, we find that for neighbouring impurities (C(1,2) and N(1,2)) the coupling energy is at least one order of magnitude larger than for the distant ones (C(1,3) and N(1,3)) which appear to be essentially decoupled. The largest energy gain is found for the case C(1,2) where our calculation shows, that in the FM case the spin-down gap almost vanishes, so that the opening up of the gap in the AFM state leads to the observed energy gain. The band gaps are reduced compared to the case of a single impurity, the strongest effect is found for C(1,2) where the interaction of the neighbouring C atoms reduces the gap to 1.4 eV. For the N substitution we find that the "distant" N(1,3) case is lower in energy, while for the C substitution the "close" C(1,2) configuration is more stable, which is again a consequence of the stronger interaction and points to a tendency of the C atoms to form clusters. The multiple nitrogen substitution configuration N(1,3) has a lower $E^{form}$ per impurity atom than the single substitutional case N(1).

\begin{table}[h]
\centering
\caption{Multiple carbon and nitrogen substitution: Cell volume, total magnetic Moment $M_{tot}$ of the supercell, defect formation energy per impurity atom $E^{form}[X]$, energy difference of the FM and AFM state $\Delta E[FM-AFM]$ and direct band gap for spin-up and spin-down.}
\renewcommand{\arraystretch}{1.15}
\begin{tabular}{lccccc} 
	X=   & C (1,2)	& N (1,2)  & C (1,3)  & N (1,3)	\\
	\hline
	Cell Volume (\AA$^{3}$)     & 522.85(+2.5\%)  & 519.51(+1.8\%)  & 522.85(+2.5\%)   & 517.87(+1.5\%) \\
	$M_{tot}$ ($\mu_{B}$)       & 4       & 2       & 4                   & 2 \\
	$E^{form}[X]$ (eV)          & 9.62    & 5.47    & 9.65                & 5.41 \\
	$\Delta E[FM-AFM]$ (eV)     & 0.160    & 0.018   & 1.14$\times10^{-3}$ & -0.019$\times10^{-3}$ \\  
 	Spin up gap (eV)            & 1.40    & 2.45    & 1.90                & 3.16 \\
 	Spin down gap (eV)          & 1.40    & 2.45    & 1.90                & 2.33 \\
\label{table:twoimp}
\end{tabular}
\end{table}

\begin{figure}
  \centering
   \subfigure[C(1,2)]{\includegraphics[width=0.45\textwidth, height=0.23\textheight]{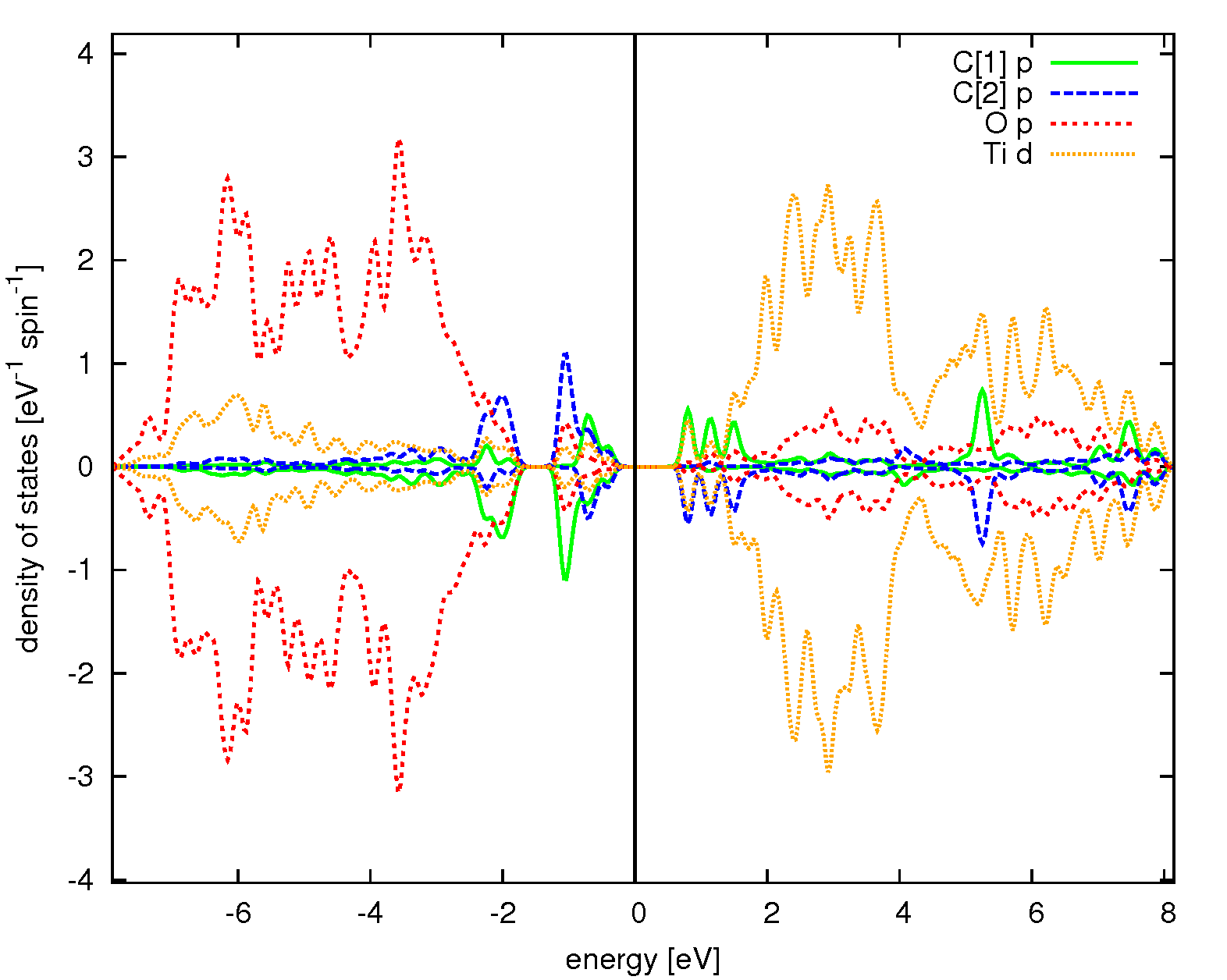}} \quad
   \subfigure[C(1,3)]{\includegraphics[width=0.45\textwidth, height=0.23\textheight]{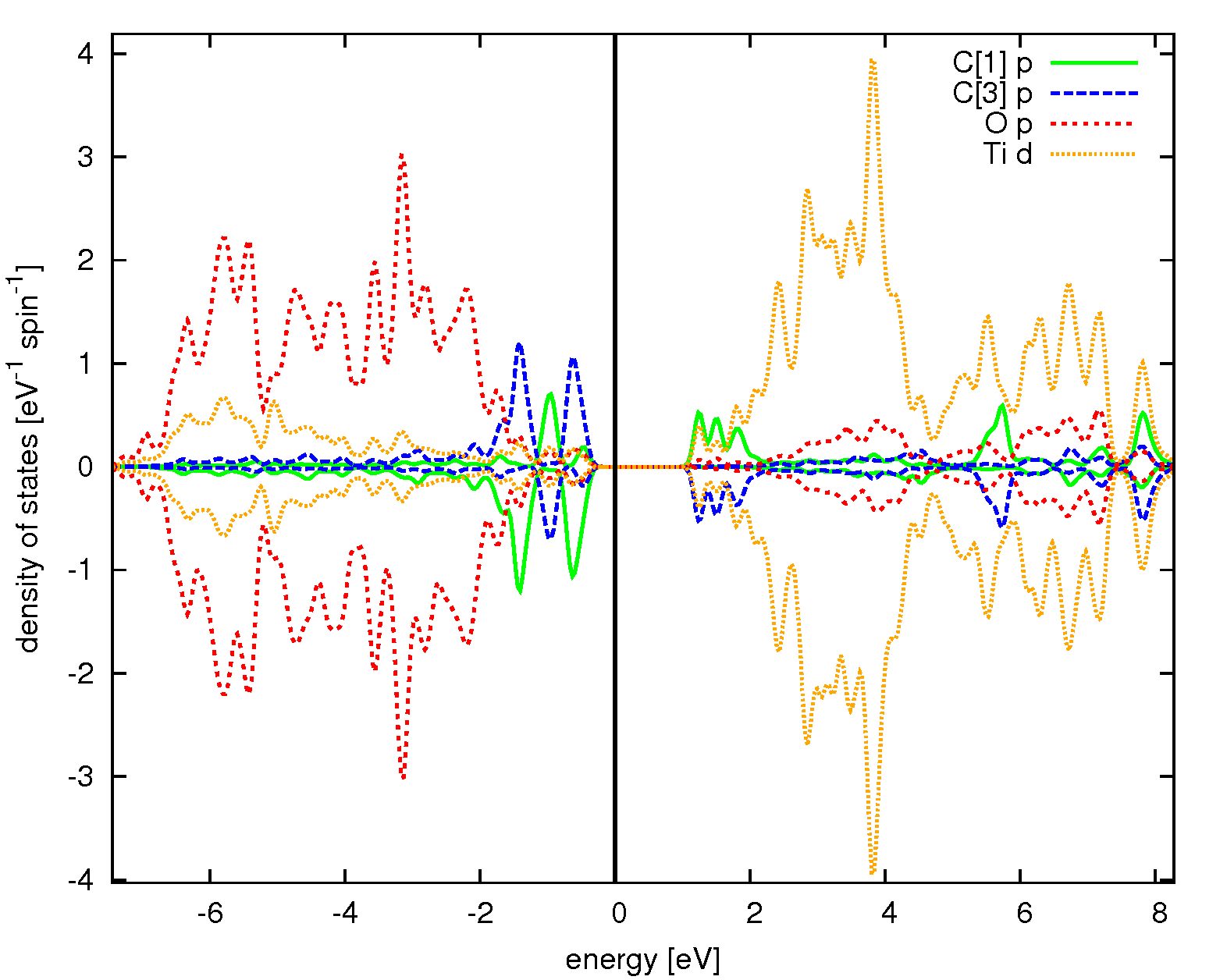}}\\
   \subfigure[N(1,2)]{\includegraphics[width=0.45\textwidth, height=0.23\textheight]{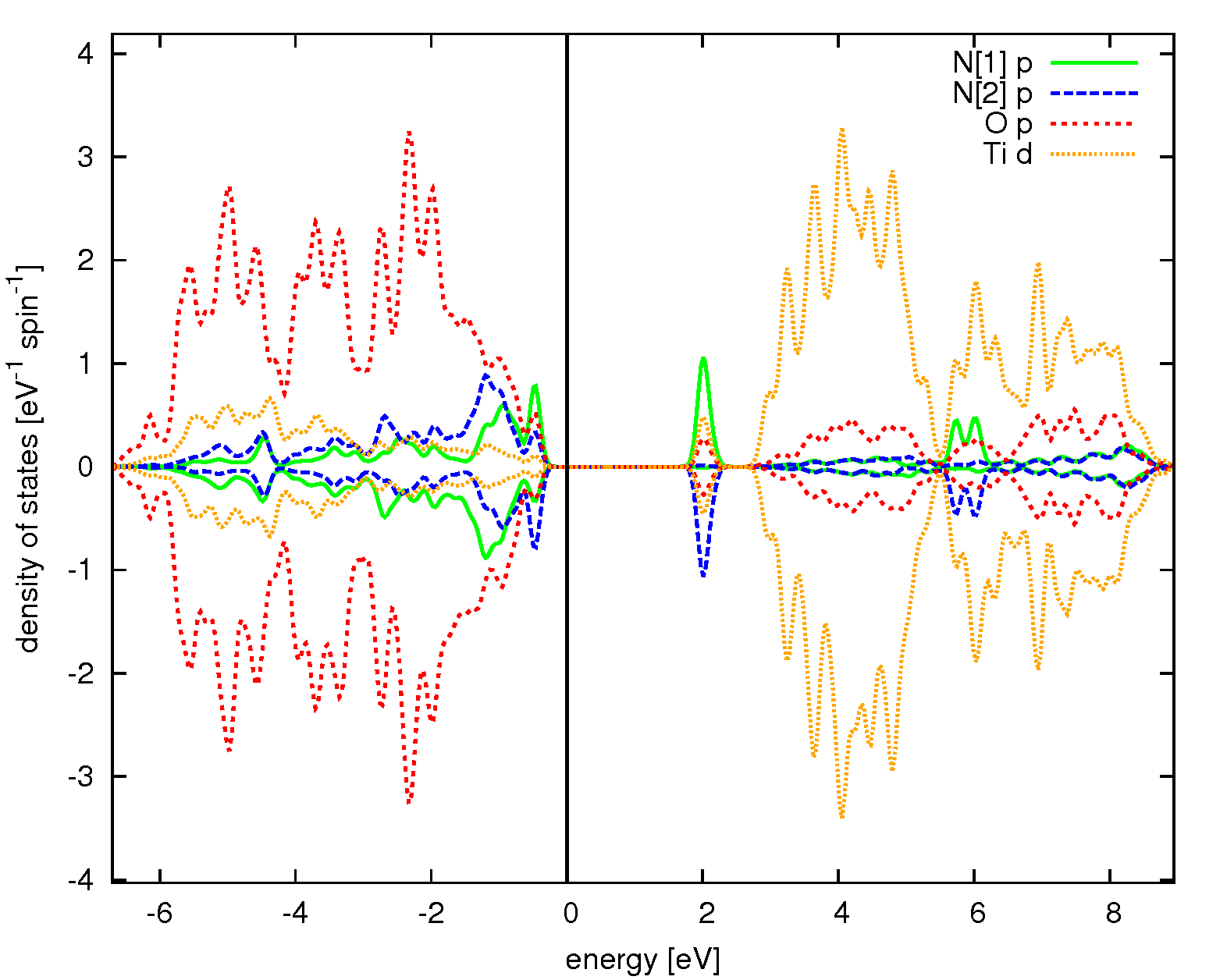}} \quad
   \subfigure[N(1,3)]{\includegraphics[width=0.45\textwidth, height=0.23\textheight]{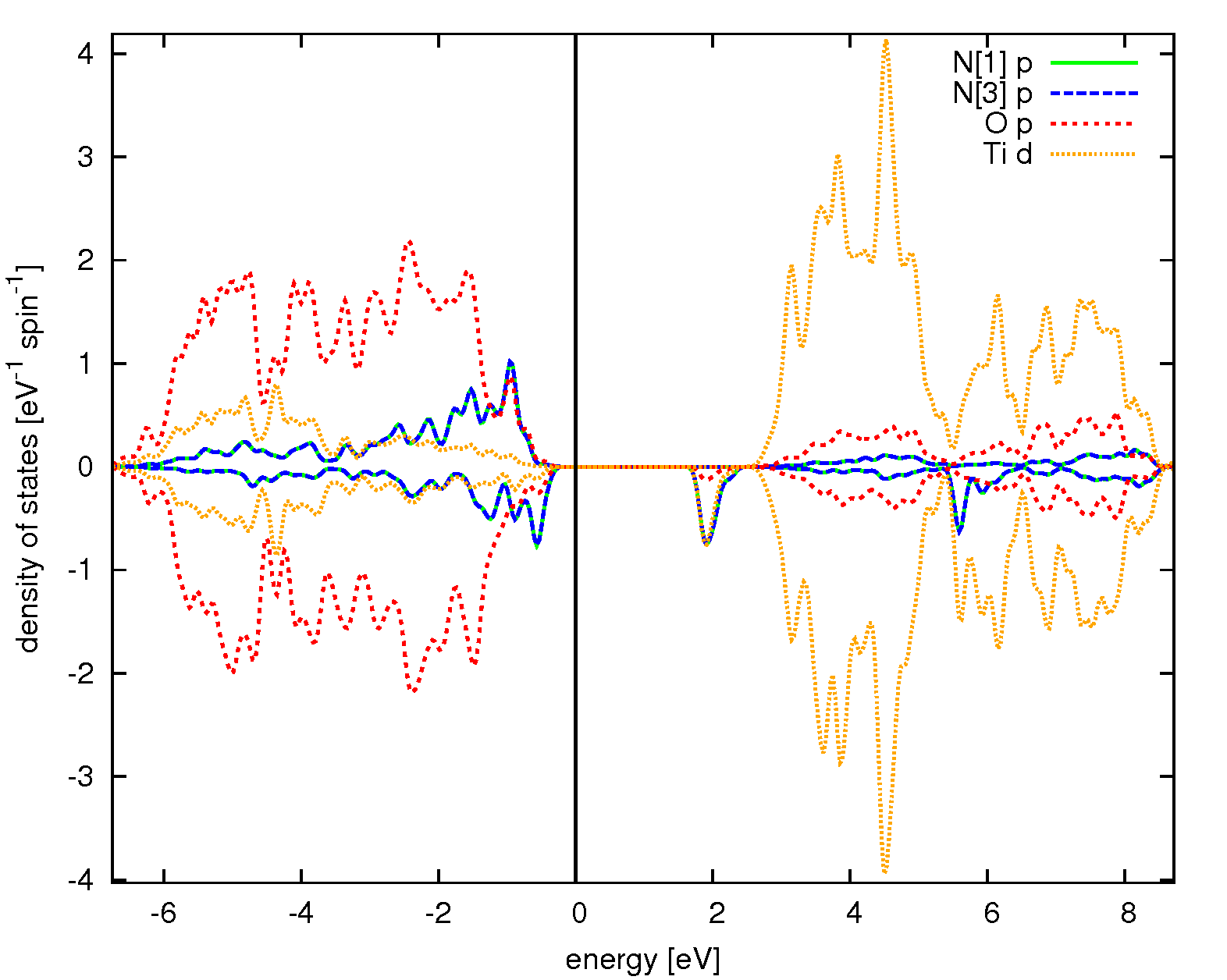}}\\
   \caption{Densities of states for the substitution of 2 oxygens by 2 carbons or 2 nitrogens. The densities of states are given for the calculated ground state which is AFM for C(1,2), C(1,3), N(1,2), and FM for N(1,3). The "bulk" oxygen and titanium states are plotted for 5 neighbouring oxygens and 2 titaniums.}
\label{fig:doppel}
\end{figure}
 
\section{C/N-interstitial}

\begin{figure}
  \centering
  \includegraphics[width=0.5\textwidth, angle=-90]{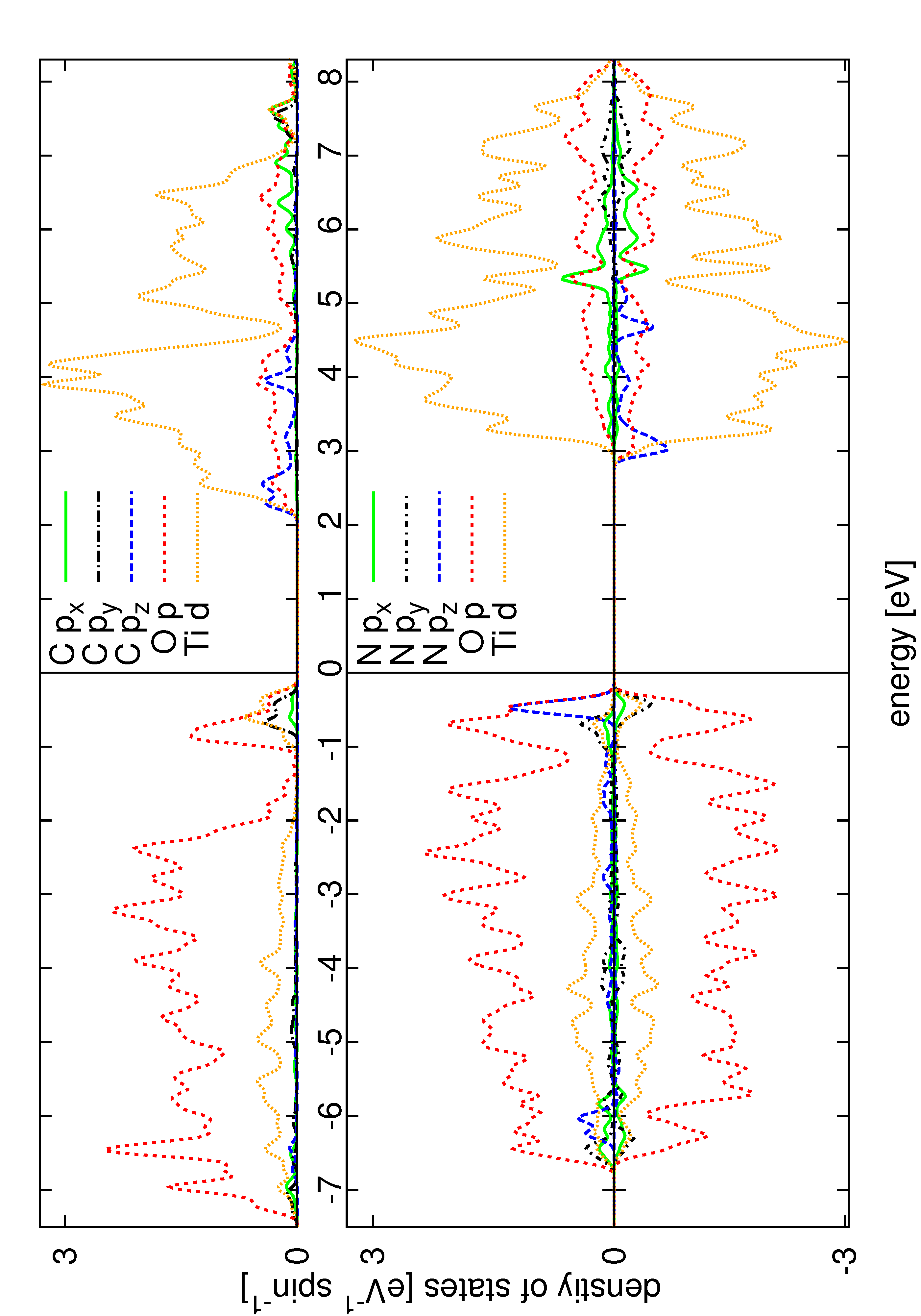}
  \caption{Densities of states for interstitial carbons or nitrogen placed on site (4).  The densities of states are given for the calculated ground state which is non-magnetic for carbon, and magnetic for nitrogen. The "bulk" oxygen and titanium states are plotted for 5 neighbouring oxygens and 2 titaniums.}
   \label{fig:interCN}
\end{figure}

\begin{figure}
  \centering
   \subfigure[C(5)]{\includegraphics[width=0.35\textwidth, height=0.23\textheight]{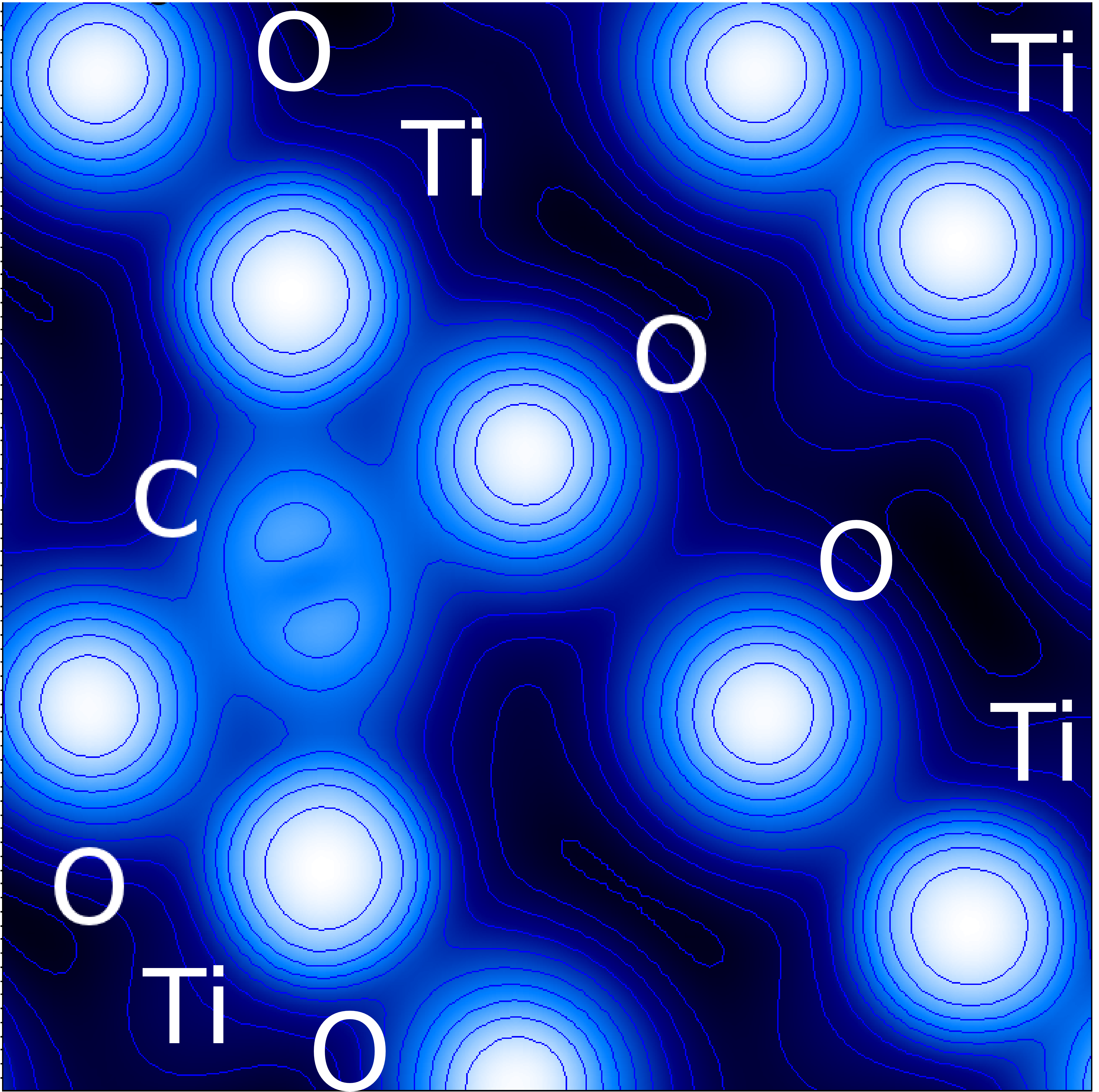}} \quad
   \subfigure[C(5)-O]{\includegraphics[width=0.35\textwidth, height=0.23\textheight]{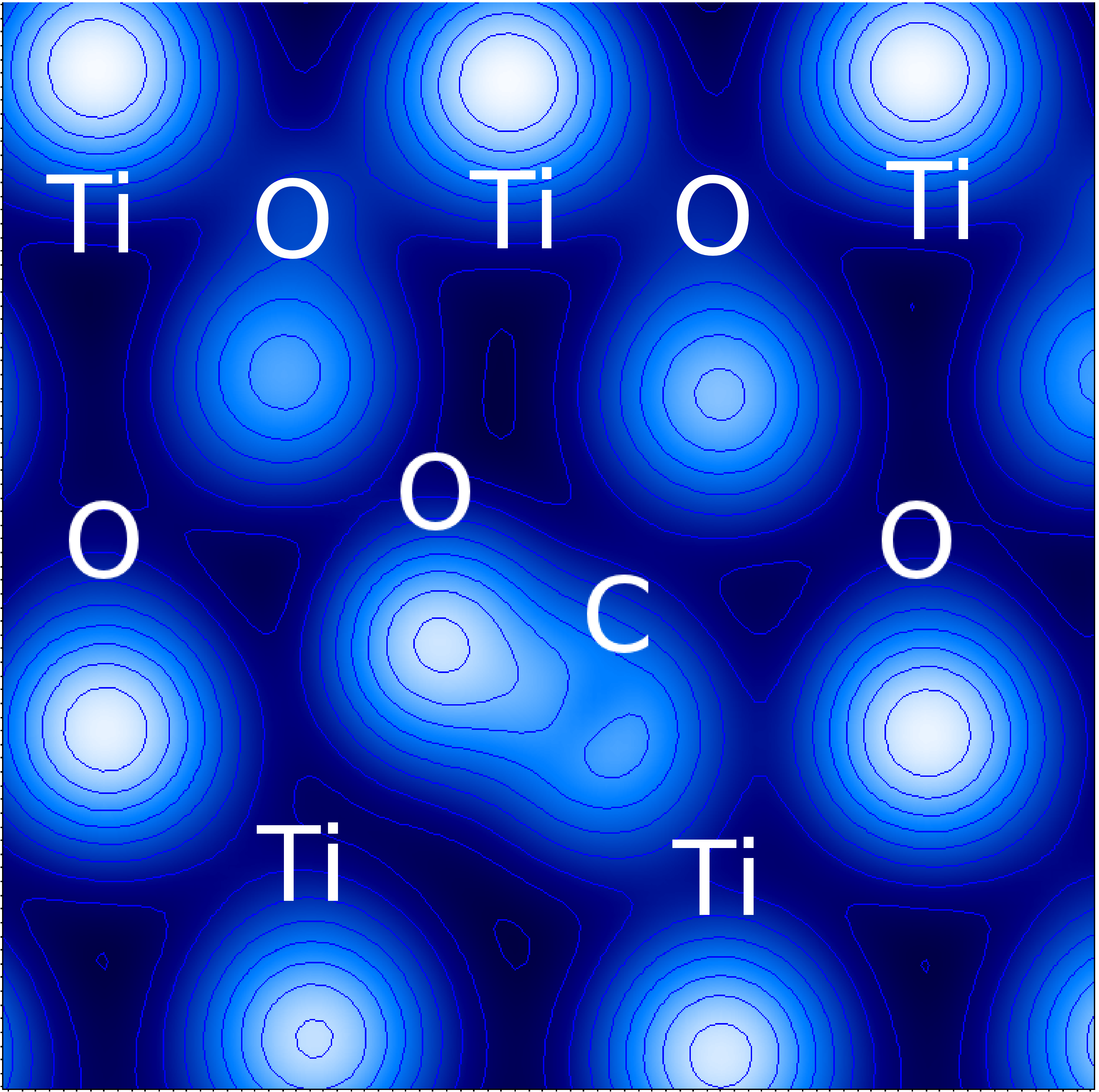}}\\
  \caption{Electron-density on a logarithmic scale for substitutional C and the CO dimer in the TiO${_2}$ lattice, lighter colors denote larger electron densities. (a) Electron-density of the non-magnetic saddle point configuration of C; (b) Electron-density of the stable configuration of CO. (b) and (a) are different planes of the crystal lattice, since the bond-axes of the CO dimer turns during formation out of plane (a).}
  \label{fig:tausch_int_CN1}
\end{figure}

\begin{figure}
  \centering
   \subfigure[C(4)-O]{\includegraphics[width=0.35\textwidth, height=0.23\textheight]{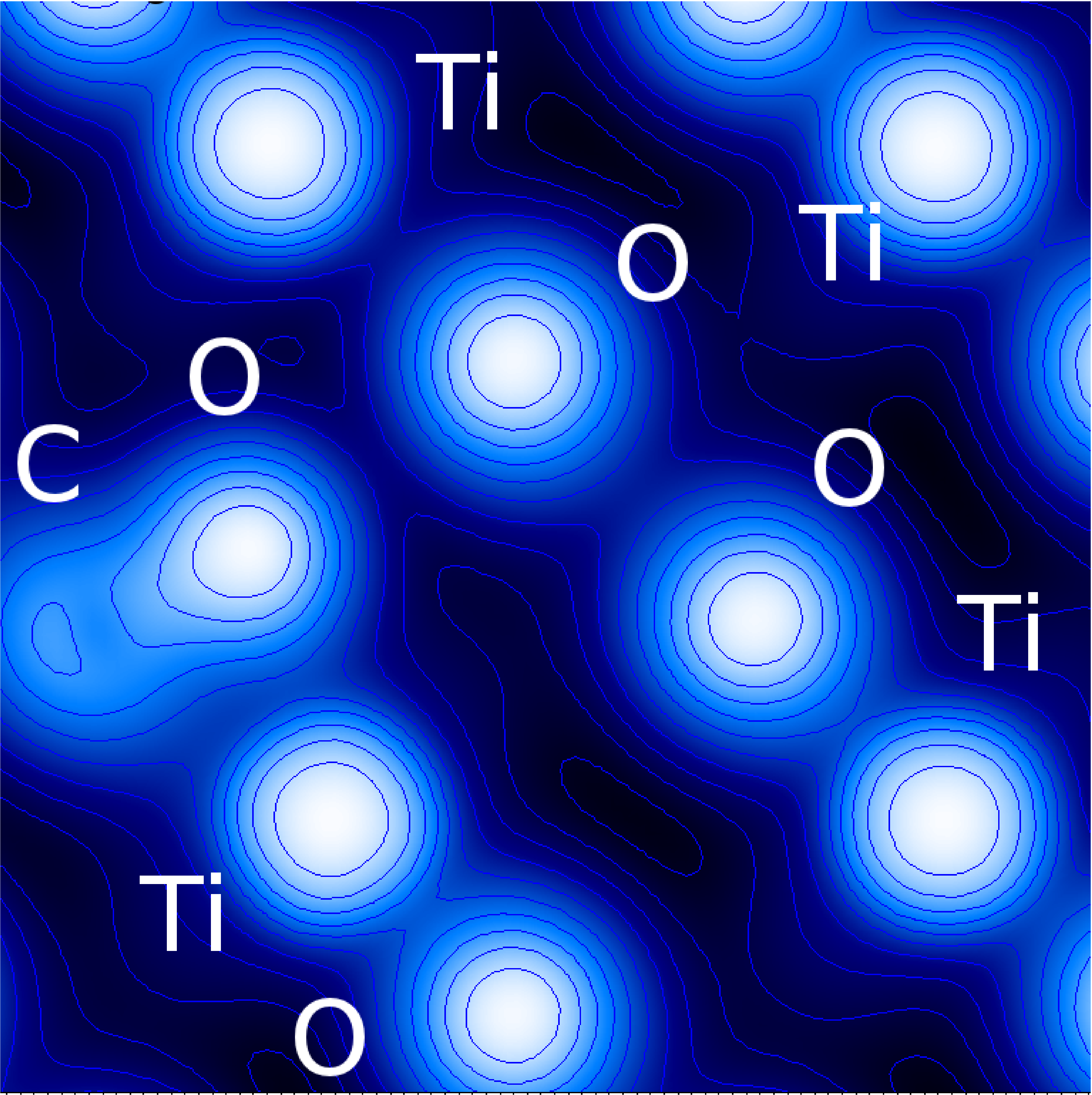}} \quad
   \subfigure[C(4)-O]{\includegraphics[width=0.35\textwidth, height=0.23\textheight]{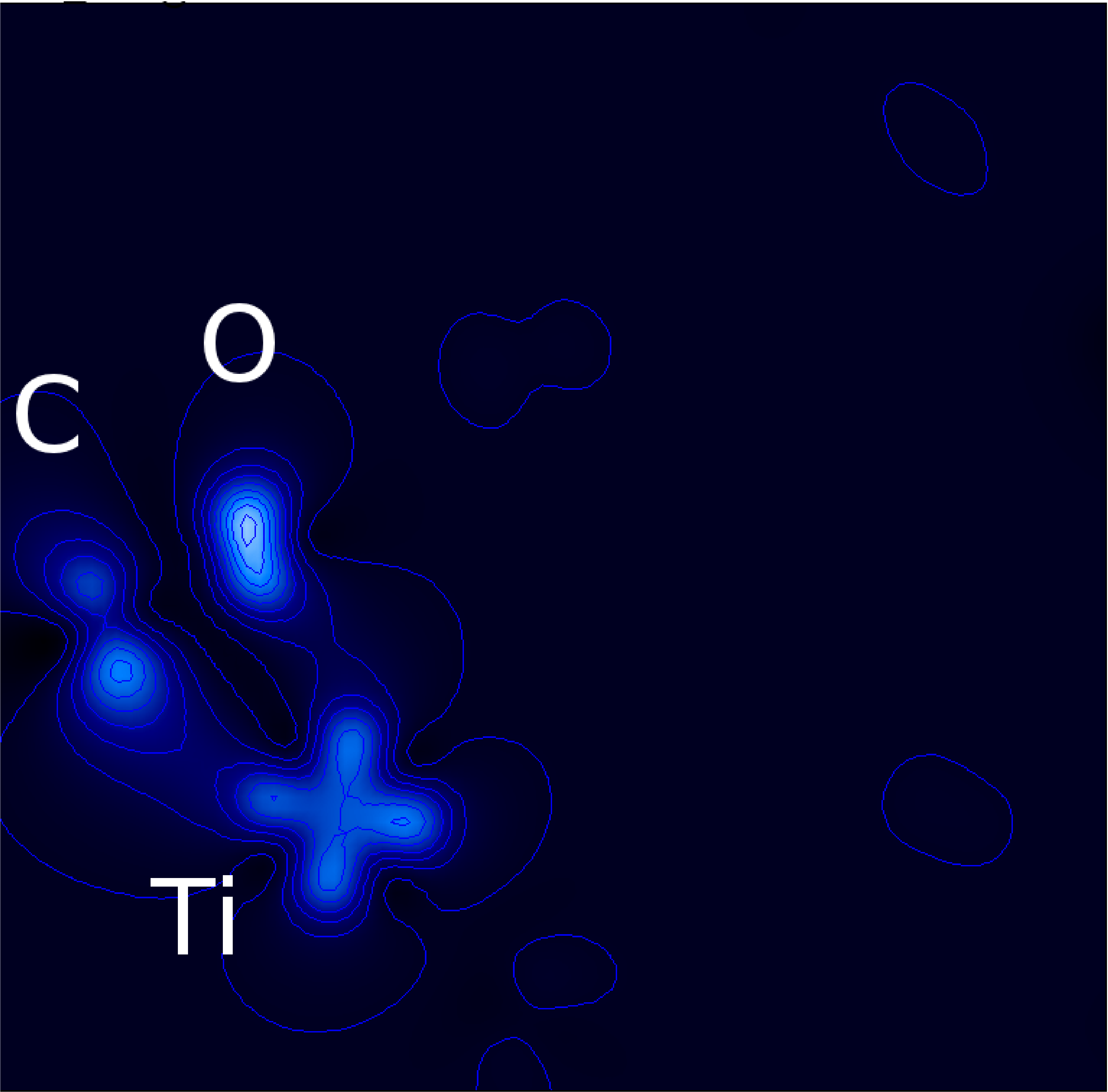}}\\
   \subfigure[N(4)-O]{\includegraphics[width=0.35\textwidth, height=0.23\textheight]{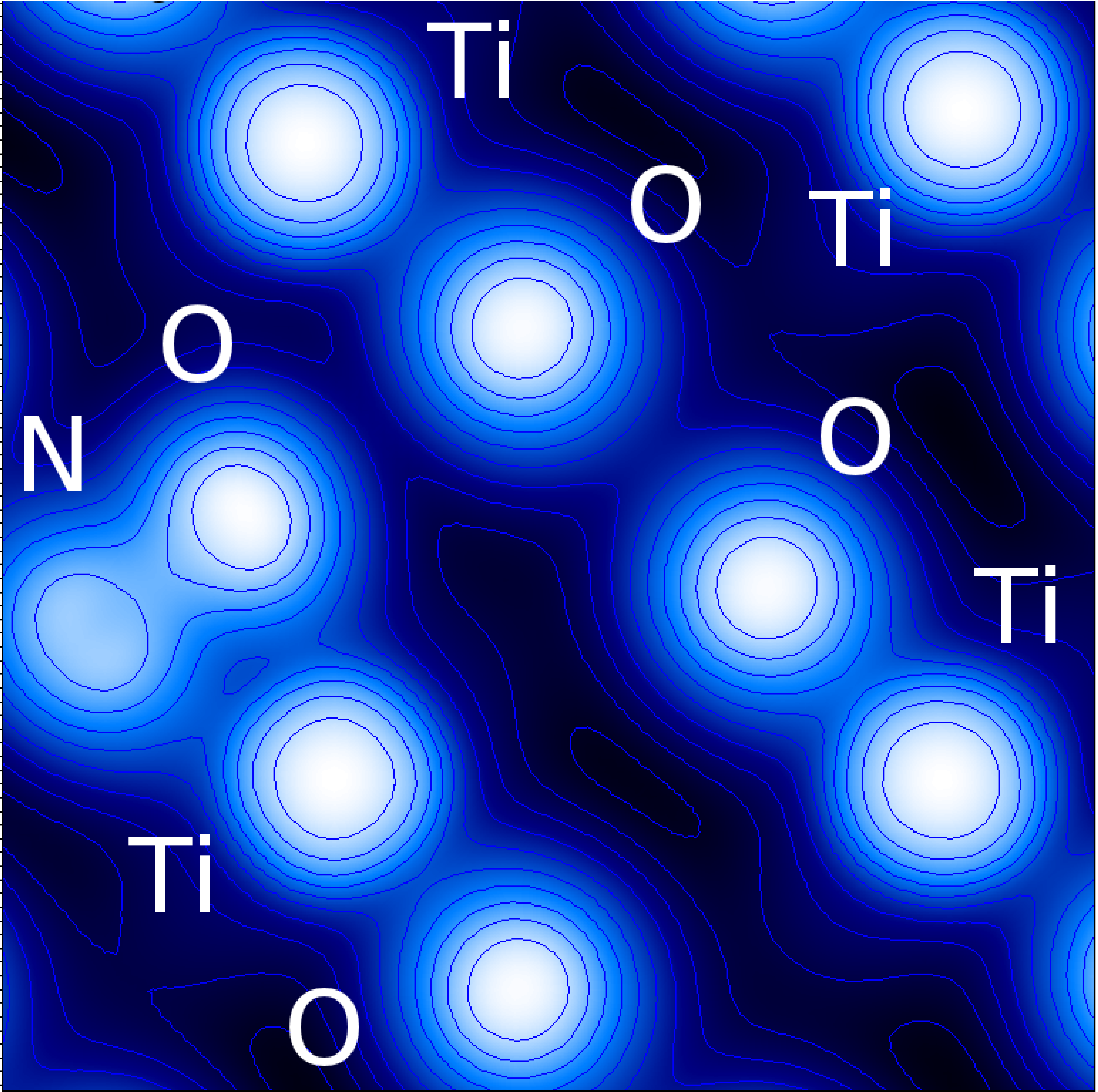}} \quad
   \subfigure[N(4)-O]{\includegraphics[width=0.35\textwidth, height=0.23\textheight]{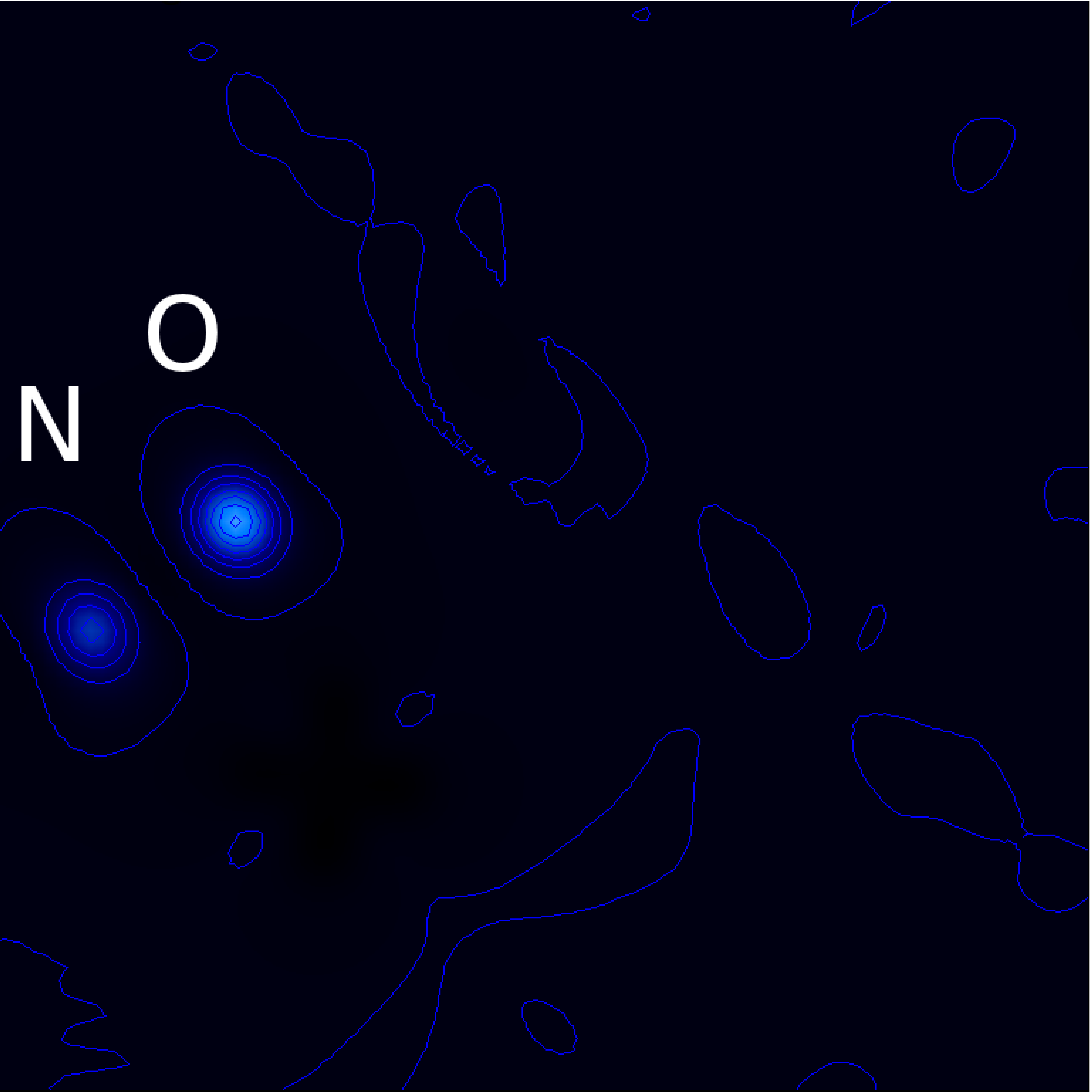}}\\
   \caption{Electron (logarithmic scale) and spin-density (linear scale) for substitutional CO and NO dimers in the TiO${_2}$ lattice, lighter colors denote larger electron densities. Electron (a) and spin density (b) of the metastable magnetic (2$\mu{_B}$) configuration of CO; Electron (c) and spin density (d) of the stable magnetic (1$\mu{_B}$) configuration of NO.}
\label{fig:tauschChg}
\end{figure}

The final investigation deals with carbon and nitrogen placed on interstitial sites (sites (4),(5) in Fig.\ref{fig:supercell}) and their interaction with the respective neighbouring oxygen. Upon putting carbon on either of the 2 interstitial sites we find that the electronic structure is very similar and no magnetic moment appears. We thus restrict ourselves to present only the density of states for interstitial site (4) (see figure \ref{fig:interCN}), the respective electron density is given in figure \ref{fig:tausch_int_CN1}(a). The interaction of the interstitial carbon with the host lattice is almost negligible and C forms atomic like flat bands at the bottom of the TiO$_{2}$ gap, which is also reflected in the electron density. The non-magnetic state is easily explained from the even number of electrons. In contrast to free carbon, which has a total spin of 1, the small but present crystal field from the host lattice leads to a double occupation of a single $p$-orbital and hence a total spin of zero. The same mechanism can be applied to nitrogen with the only difference that N has 3 $p$-electrons and consequently always one singly occupied orbital with a resulting total spin of $\frac{1}{2}$. Energetically we find that C/N sits on a saddle point of the total energy surface. Upon shifting C/N slightly to either side, the interstitial atoms approach their oxygen neighbours and form CO and NO molecule type entities which is accompanied by a dramatic change in the electronic structure. The highly reactive oxygen tries to reach its 2$^-$ state, by forming a CO or NO molecule which ends up almost at the original oxygen position. The electron density once the C(5)-O dimer formed is shown in figure \ref{fig:tausch_int_CN1}(b).  We also performed a second set of calculations, where C/N are placed on an oxygen position and this oxygen is put at an interstitial site. Starting from this configuration and after relaxation again led to the formation of CO and NO dimers with the same total energy as before. The only difference is that the C/N interstitial sits at an energy saddle point, while the oxygen interstitial immediately starts to move towards the C/N. Fig.\ref{fig:tauschChg}(a) and (c) shows the charge densities for the CO and NO case. One clearly sees the molecular structure with a bond length of 1.22 \AA\ for CO and 1.31 \AA\ for NO. For carbon the bond length agrees with the CO double-bond length of 1.20 \AA , for nitrogen the agreement is less good and lies between the NO single-bond length of 1.45 \AA\ and the double-bond length of 1.17 \AA\ \cite{Marye1994} which may be caused by the fact that the NO "molecule" exhibits a magnetic moment of 1$\mu_{B}$.

Table \ref{table:int_tausch} summarizes the results of our calculations. The first 2 lines contain the data for the C/N interstitials at their saddle point position. Also for this non-equilibrium state a semiconductor is found. Again nitrogen becomes magnetic with 1$\mu_B$, while carbon remains non-magnetic. The stable configuration is found, when both carbon and nitrogen become inserted on site (5). For both cases a CO or NO dimer is found, which considerably lowers the total energy. In an experimental study done by Chen et al.\cite{ChenBurda2005} nitrogen doped TiO$_2$ nanocolloids have been prepared and their photocatalytic properties were investigated. XPS measurements done by this group suggest the formation of NO binding regions in these samples, which again underlines our findings of the NO dimers.  Asahi et al. \cite{Asahi2007} report about experiments supported by \textit{ab-initio} calculations where they study various kinds of N complex species incorporated into a TiO$_2$ anatase host matrix and as well encountered the formation of NO bonds. DiValentin et al. \cite{DiValentin2005} again found single bonded CO dimers from ab-initio calculations. From our calculations we suggest the formation of NO and CO dimers in TiO$_2$, a possible experimental proof could be the detection of the (symmetric) stretching mode frequency, which we calculate to be 1605cm$^{-1}$ for CO and 1234cm$^{-1}$ for NO inside the TiO$_2$ lattice.   

The defect formation energies are lower for the interstitial configuration compared to the oxygen substitutional case. For N(5) $E^{form}$ decreases 1.5 eV compared to the single substitutional case N(1) and $E^{form}$ for interstitial C(5) and substitutional C(1) even differ by 4.35 eV.  

\begin{table}[ht]
\centering
\caption{C/N-interstitial: Atomic distance of C/N and its nearest neighbouring oxygen, total magnetic Moment $M_{tot}$ of the supercell, defect formation energy per impurity atom $E^{form}[X]$, and direct band gap for spin-up and spin-down.}
\scalebox{0.80}{
\renewcommand{\arraystretch}{1.20}
\begin{tabular}{lcccccccc} 
	                               & C(4)         & N(4)          &   & C(4)-O       & C(5)-O       & N(4)-O      & N(5)-O     \\
	\hline
	Atomic Distance C/N-O (\AA)    & 1.80(+6.4\%) & 1.81(+6.9\%)  &   & 1.22         & 1.24         & 1.32        & 1.32       \\
	$M_{tot}$ ($\mu_{B}$)          & 0            & 1             &   & 2            & 0            & 1           & 1          \\ 
	$E^{form}[X]$ (eV)             & 6.92         & 5.33          &   & 6.04         & 5.05         & 4.08        & 3.91      \\
 	Spin up gap (eV)               & 2.59         & 3.51          &   & 0.77         & 1.92         & 2.89        & 3.16       \\
    Spin down gap (eV)             & 2.59         & 3.40          &   & 3.39         & 1.92         & 2.32        & 2.35       \\
\label{table:int_tausch}
\end{tabular}}
\end{table}

\begin{figure}
  \centering
  \includegraphics[width=0.5\textwidth, angle=-90]{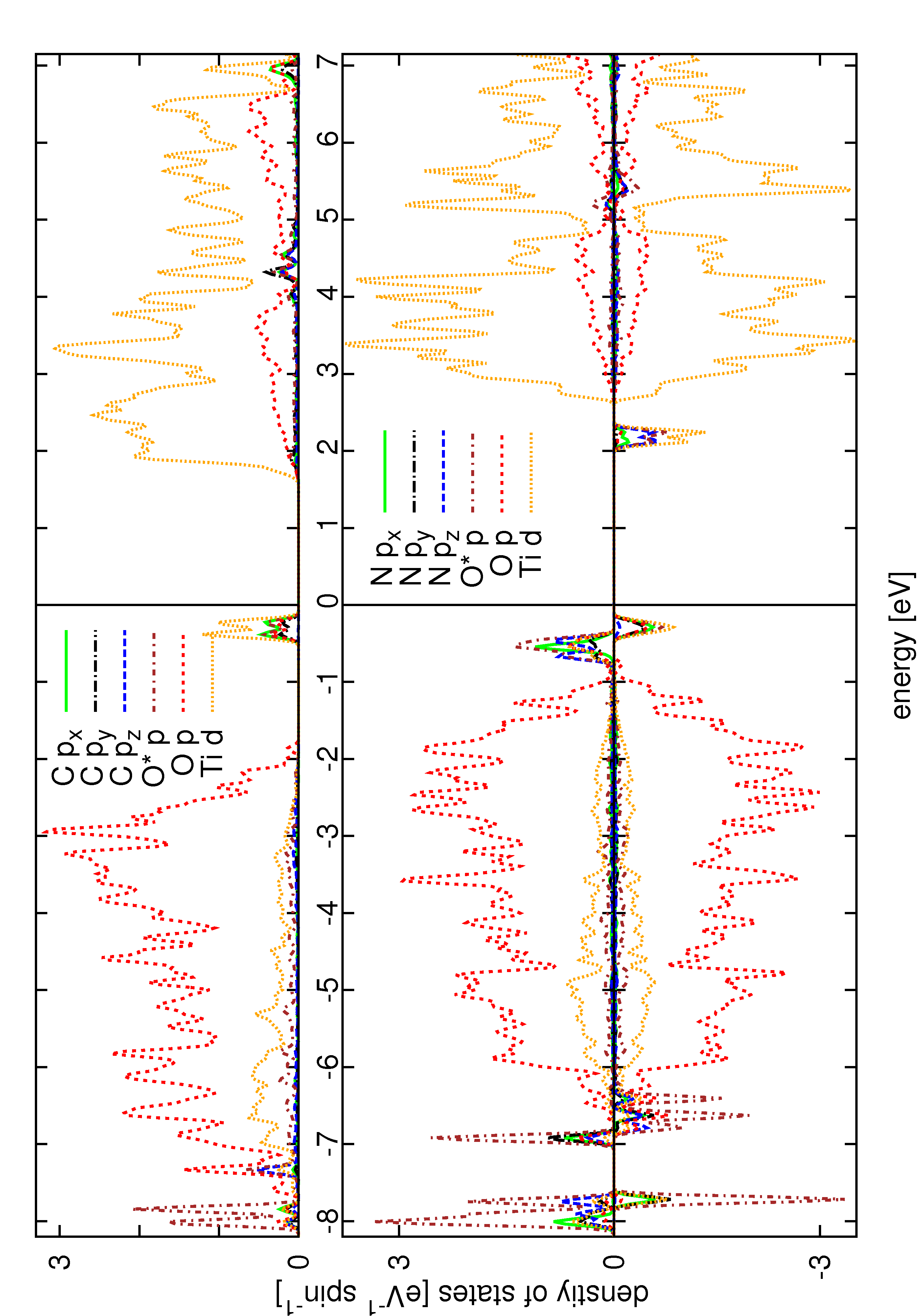}
  \caption{Densities of states for the interstitial C(5)-O and N(5)-O doping configuration. The densities of states are given for the non-magnetic state for carbon, and the magnetic state for nitrogen. The "bulk" oxygen and titanium states are plotted for 5 neighbouring oxygens and 2 titaniums. O$^*$ is the oxygen of the CO or NO dimer.}
  \label{fig:tausch_CN3}
\end{figure}

The densities of states (Fig.\ref{fig:tausch_CN3}) show the basically different behaviour as compared to the C/N impurity on the saddle point state (see Fig.\ref{fig:interCN}). The oxygen states are below the bulk oxygens and the C/N $p$-states are inside the gap. The charge transfer of 2 electrons leads to a strong splitting of the occupied oxygen and the unoccupied carbon/nitrogen states. It is noteworthy that there exists also an interaction of the high lying carbon $p$-states and the Ti $d$-states. The NO interaction is weaker, which is established by the larger bond length. The oxygen states are again below the oxygen bulk, but the nitrogen $p$-states are just at the bottom of the TiO$_2$ gap. The unpaired electron leads to a magnetic moment of 1$\mu_B$ and the interaction with the Ti $d$-states is much smaller than for the carbon case. The respective charge and spin density is shown in Fig.\ref{fig:tauschChg}(c), (d). 

For the case of carbon with oxygen on position (4) we find a metastable magnetic solution with a magnetic moment of 2$\mu_B$ per supercell. Unlike the cases discussed earlier in this paper, the magnetic moment is no longer located at carbon only, but is distributed between carbon, oxygen and 3 neighbouring Ti atoms. Figure \ref{fig:tauschChg}(a), (b) shows the electron density and the spin density, respectively, where the polarization of the Ti $d$-states is easy to recognize. However, due to the unfavourable total energy of this magnetic state we expect that this case will not be present in real samples.

\section{Discussion}

\begin{figure}
  \centering
   \subfigure[Phase diagram $\mu_{C}/\mu_{O'}$]{\includegraphics[height=0.20\textheight]{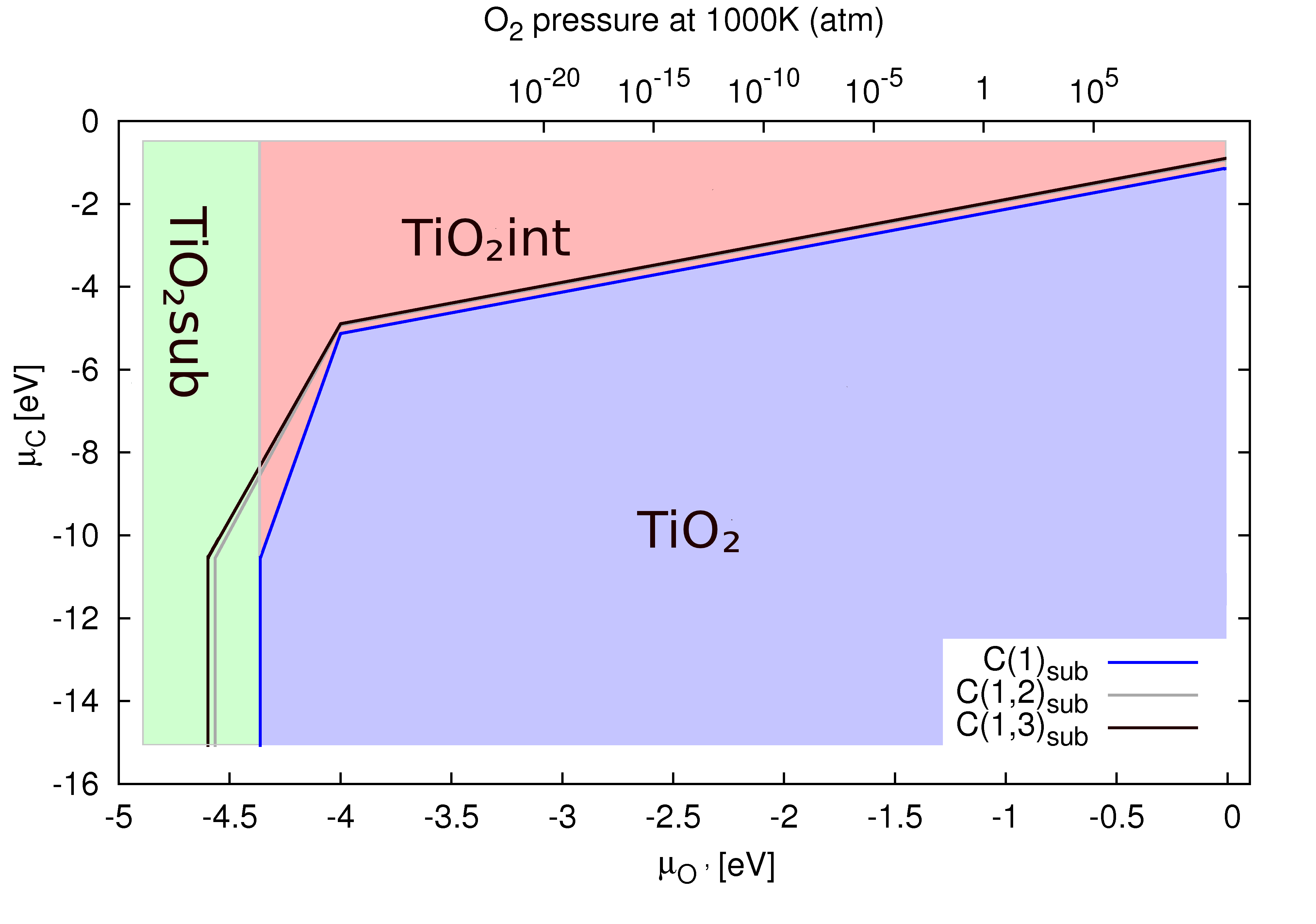}} \quad
   \subfigure[C-doped $E^{f}$/$\mu_{O'}$]{\includegraphics[height=0.20\textheight]{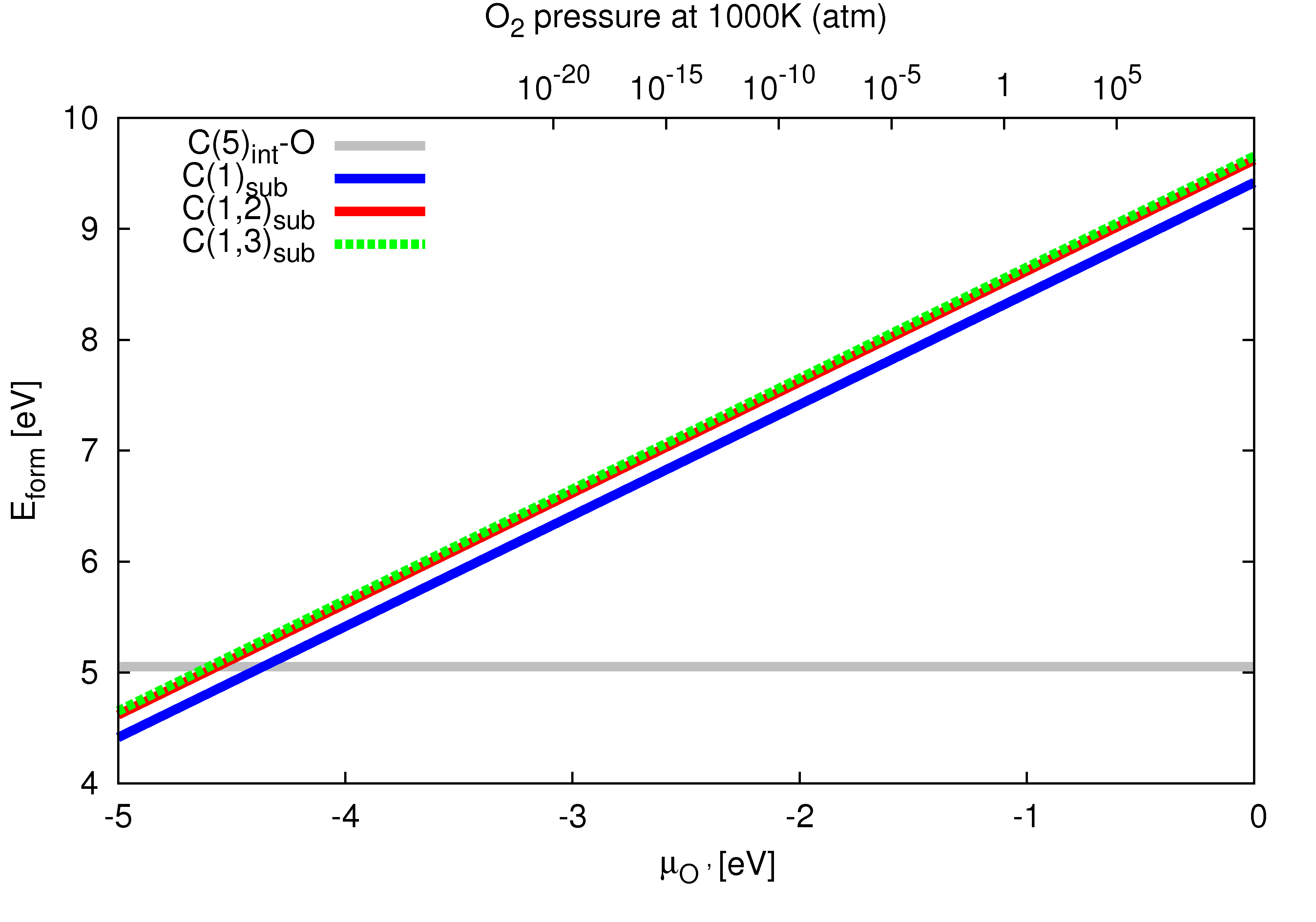}} \\
   \subfigure[Phase diagram $\mu_{N}/\mu_{O'}$]{\includegraphics[height=0.20\textheight]{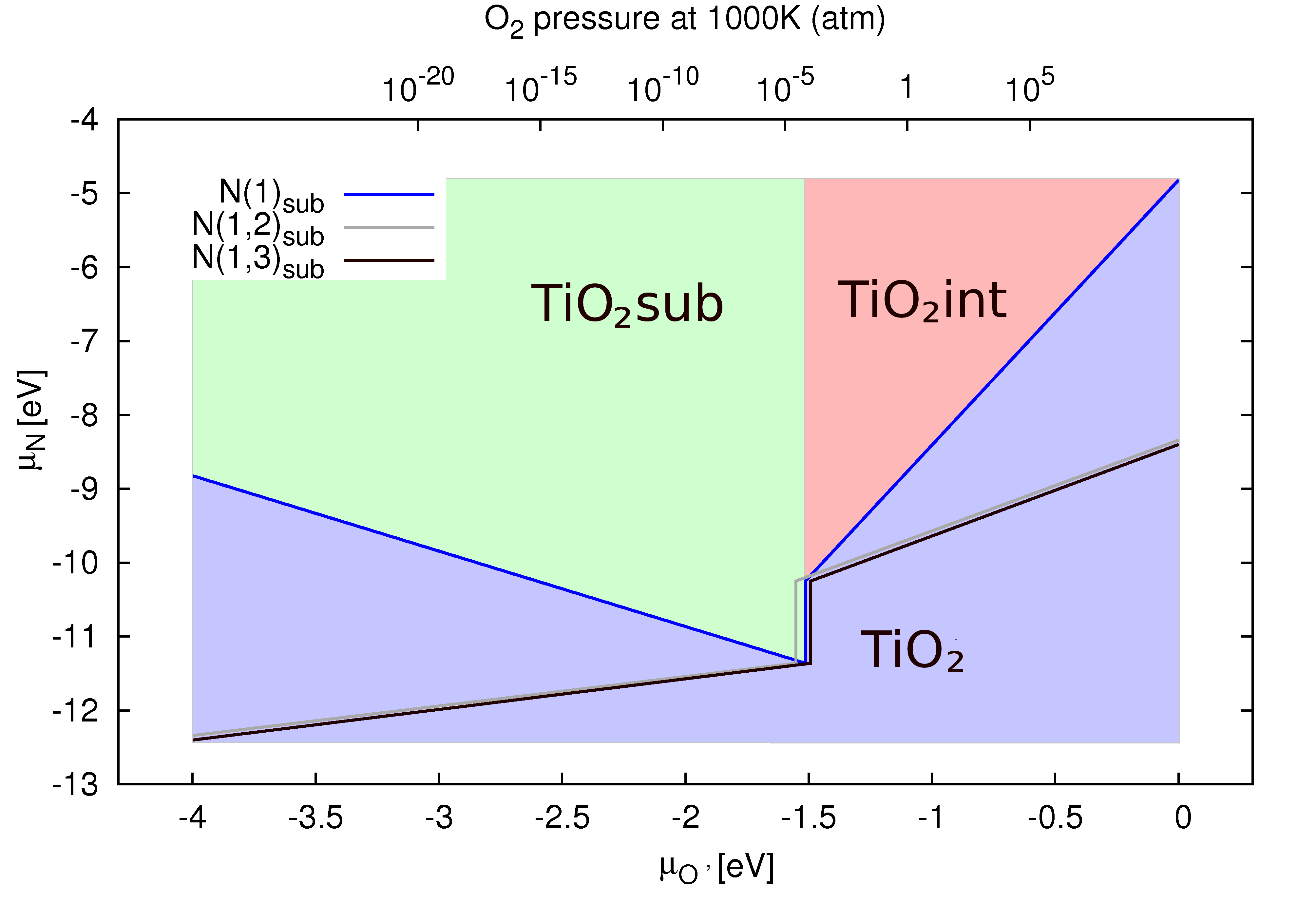}} \quad
   \subfigure[N-doped $E^{f}$/$\mu_{O'}$]{\includegraphics[height=0.20\textheight]{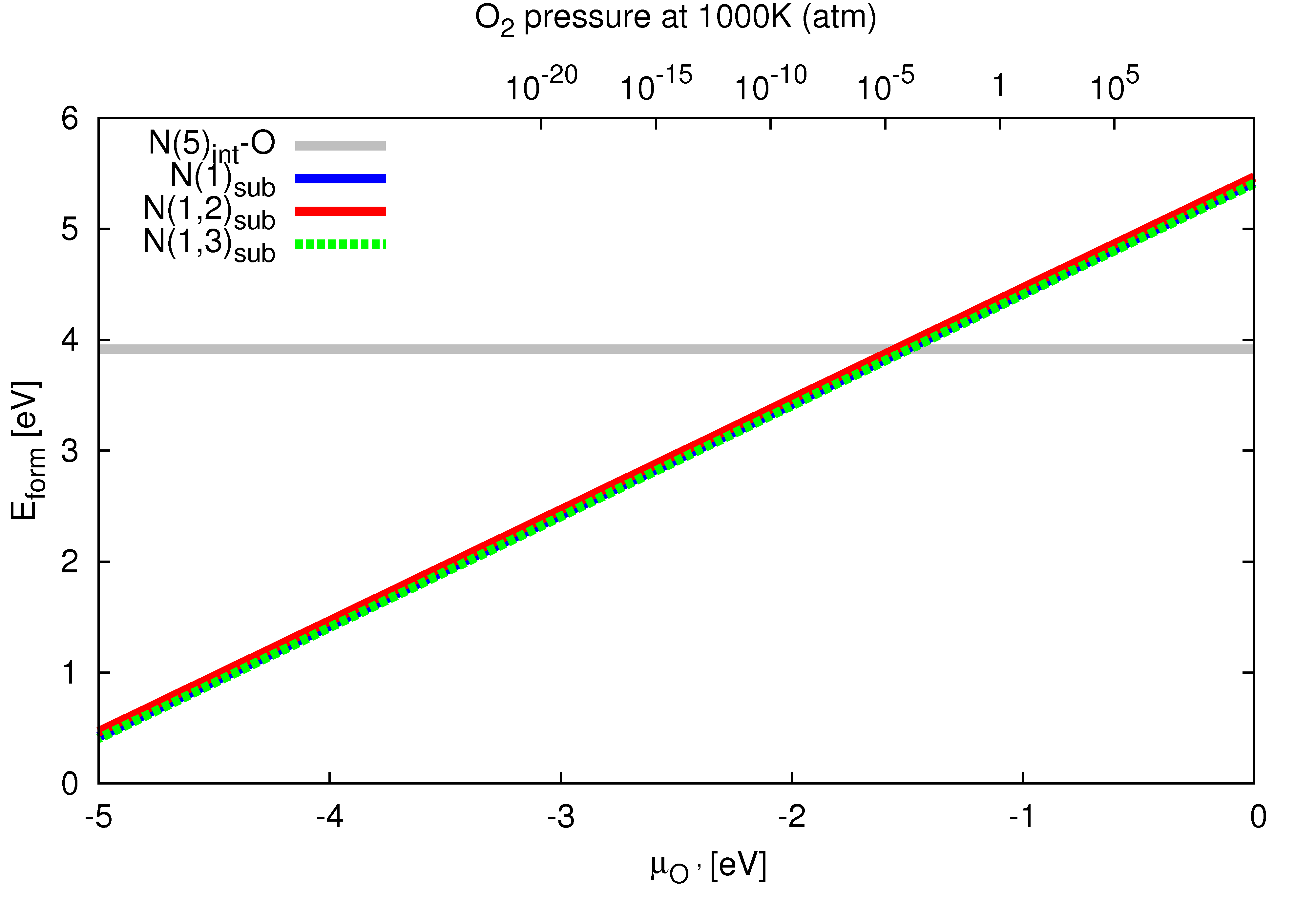}}\\
   \caption{(color online)(a) and (c): phase diagrams as a function of the chemical potentials $\mu_{O'}$,$\mu_{C}$ and $\mu_{N}$; (b) and (d): formation energies $E^{form}[X]$ as a function of the oxygen chemical potential $\mu_{O'}$.}
\label{fig:phasediagram}
\end{figure}

Fig. \ref{fig:phasediagram} shows the calculated $\mu_{C}$/$\mu_{O'}$ and $\mu_{N}$/$\mu_{O'}$ phase diagrams as well as the corresponding defect formation energy $E^{form}[X]$ versus oxygen chemical potential $\mu_{O'}$. In Fig.\ref{fig:phasediagram}(a) $\mu_{C}$ vs. $\mu_{O'}$ is plotted giving the thermodynamically stable phases of carbon doped TiO$_{2}$ dependent on the chemical potentials of C and O. On the top x-axis the oxygen chemical potential is converted into equivalent oxygen pressure at a fixed temperature (T=1000K). Highlighted in color are the thermodynamically different phases for the single carbon substitutional case C(1). The phase boundaries for the multiple substitutional cases C(1,2) and C(1,3) are indicated by grey and black lines, respectively. Comparing these three phase boundaries shows that they are almost identical over a large range of $\mu_{O'}$ and only differ for very small chemical potentials of oxygen. 
For a large range of $\mu_{O'}$ and $\mu_{C}$ the pure undoped TiO$_{2}$ is the thermodynamically preferred phase (Fig.\ref{fig:phasediagram}(a)). Only for very small oxygen chemical potentials, i.e. an oxygen poor phase, the substitution of oxygen by carbon becomes favourable. Interstitial doping of rutile is energetically practicable only for higher carbon chemical potentials together with an oxygen rich environment. Different values of $\mu_{C}$ correspond to different reservoirs of carbon atoms. $\mu_{C}=\frac{1}{2}\mu_{C_{Diamond}}$=-10.55 eV corresponds to a reservoir of C in Diamond and $\mu_{C}=\mu_{CO_{2}}-\mu_{O2}$=-15.17 eV to C in CO$_2$. That indicates that for interstitial carbon doping using CO$_2$ as a C reservoir only the substitutional and the pristine crystal phases are stable configurations.
Fig. \ref{fig:phasediagram}(b) shows the intersection points of the defect formation energy for the different doping configurations C(1), C(1,2), C(1,3). As a reference for the carbon chemical potential Diamond was chosen as a reservoir. The grey line is the interstitial carbon doped case C(5) which has a constant defect energy and does not vary with the oxygen chemical potential. Only for $\mu_{O'}$ about $\leq -4.5$eV the substitutional phases become thermodynamically more stable. Since all intersection points for the different doping configurations are located close to each other at extremely low oxygen pressure, sole thermodynamical reasoning would only allow for interstitial doping.

Fig. \ref{fig:phasediagram}(c) shows the $\mu_{N}$ over $\mu_{O'}$ phase diagram. In contrast to Fig. \ref{fig:phasediagram}(a) the thermodynamically stable are of pure TiO$_{2}$ is smaller than in the carbon case.
Thus, looking at the corresponding oxygen pressure and at the  nitrogen chemical potential, all three doping configurations become feasible. Nitrogen atoms taken out of a NO$_2$ reservoir correspond to a chemical potential of $\mu_{N}=\mu_{NO_{2}}-\mu_{O2}$=-11.36 eV and a N$_2$ reservoir leads to a nitrogen chemical potential of $\mu_{N}=\frac{1}{2}\mu_{N_{2}}$=-10.249 eV.
Again, as in the CO$_2$ case, NO$_2$ would only allow for a stable substitutional and pure phase at varying oxygen chemical potential.
Fig.\ref{fig:phasediagram}(d) shows the intersection of the interstitial and substitutional doped defect formation energy lines, where N$_2$ is assumed as the nitrogen reservoir. In contrast to the carbon case, for nitrogen doping the intersection point are in a region of realistic oxygen pressures, which may allow for both substitutional and interstitial doping.

An efficient photocatalyst is supposed to have a "clean" band gap with no additional states inside that can act as electron-hole traps and recombination centers and therefore reduce the lifetime of the photo-induced charged carries, leading to a decrease of photocatalytic efficiency. By doping our system with C and N atoms we try to narrow the band gap in an efficient way without introducing mid-band gap states which cause the recombination rate of the charge carriers to become maximum, as derived by Shockley, Read and Hall \cite{SHR1952,Hall1952}. Interpreting our calculated DOS and keeping the SRH model in mind, we see that even if there is a reduction of the band gap there can be decreased photocatalytic efficiency.    

Doping the TiO$_2$ rutile bulk structure with nitrogen atoms does not lead to the creation of impurity states at the mid-band energy level ($E_{MB}$ - defined as half of the energy difference of the top energy level of the VB and the bottom energy level of the CB). In the cases studied here, the dopant levels created inside the band gap are mostly located in the lower or top third of the band gap. Hence the life time of the charge carriers is reduced but does not reach its minimum like for the case of impurity states at $E_{MB}$. Comparing the different nitrogen doped systems, one sees that the spin-up gap remains almost unchanged, showing a maximum reduction of 0.08 eV. The spin-down gap on the other hand shows a minimal reduction of up to 0.91 eV. 
For the case of single nitrogen doping there is a negligible small narrowing of the spin-up gap by 0.04 eV. The spin-down gap is reduced by a dopant state in the upper third of the gap by 0.64 eV. For the case of two nitrogen impurities the spin-down gap again is reduced by 0.79 eV for the impurity atoms located at (1) and (2) and by 0.91 eV at (1) and (3). For the first case the spin-up gap again is reduced by 0.79 eV and for the latter one narrowed by 0.08 eV. In both systems there is, like in the case of single nitrogen doping, an impurity level introduced in the upper third of the band gap near the CB. Regarding the unstable interstitial state we see a blueshift since both gaps are widened. The spin-up gap by 0.27 eV and the spin-down gap by 0.16 eV. For the case of the NO bond like formation the band gap narrowing for the spin-up state again is inconsiderably with a reduction of 0.08 eV. The spin-down gap is reduced by 0.89 eV. Impurity states are introduced in the upper and lower third of the band gap. In total nitrogen doped TiO$_2$ causes a redshift in the absorption and introduces band gap states that do not cause the recombination rate to maximize and therefore conserve a reasonable photocatalytic efficiency. The only case where a blueshift could be detected was in the unstable case of interstitial doping for nitrogen at the positions (4) and (5) without the formation of NO dimers. Hence, there have to be other reasons leading to the blueshift observations in N-doped rutile crystals.     

Doping with carbon always leads to a significant redshift but introduces mid-band gap levels at $E_{MB}$, thus maximizing the recombination rate of the electron-hole pairs causing minimal photocatalytic efficiency. For the case of single carbon doping the spin-up band reduces by 1.04 eV and the spin-down gap by 1.18 eV. New states are created inside the band gap but not in the region of $E_{MB}$. For the case of two carbon dopants at the positions (1) and (2) mid-band gaps are created and the gaps for both spin directions are reduced to 1.40 eV which is beyond the visible spectrum, rather unfavourably for photocatalysis. Carbon at the substitutional positions (1) and (3) in contrast shows a desirable behaviour with no impurity states in the region of $E_{MB}$ and a band gap of 1.90 for both spin states. Carbon on the interstitial position (4) and (5) is similar to the latter case, there are no states in the mid-band gap and a gap reduction to 2.59 eV. In the case of the CO dimer there are again states at the energy level $E_{MB}$, making the system unattractive for photocatalytic application even with a putative band gap of 1.92 eV. In summary there are only two favourable configurations of carbon in the rutile TiO$_2$ lattice, the single carbon impurity and the one with the carbon impurities at positions (1) and (3).
When we replace oxygen by C/N both impurities become magnetic. For two impurities on oxygen sites we find an anti parallel orientation of the magnetic moments to be favoured, however, the low energy gain due to the formation of magnetic order suggests that the ordering temperatures will be rather low. When C/N is placed on an interstitial site, C remains non-magnetic while N always shows a magnetic moment of 1$\mu{_B}$. Once the CO and NO dimers are formed, nitrogen remains with 1$\mu{_B}$ while carbon is non-magnetic. Test  \cite{LIU2011,QIN2014,Drera2010,BUZBY2006,WANI2013}.

\section{Conclusion}
The magnetic structure of carbon and nitrogen doped rutile TiO$_2$ was investigated for different doping configurations and concentrations. 
Carbon shows to induce a magnetic moment of 2$\mu_{B}$ per unit cell for the substitutional doping case and a nonmagnetic solution for the interstitial doping configuration where a CO dimer is formed. For both configurations the bandgap is reduced, whereas for the CO dimer case midband states are created which could lead to a reduction in the photocatalytic performance. Two substitutional C impurities close to each other show AFM coupling and for the case of maximum distance paramagnetic behaviour. Relating to the phase diagram carbon is most stable as an interstitial impurity for low chemical potentials.
Nitrogen induces a magnetic moment of 1$\mu_{B}$ for both the substitutional and the interstitial doping configuration. For latter a NO dimer is formed like in the interstitial Carbon doping case. Nitrogen doping reduces the bandgap of rutile TiO$_2$ as well, whereas no midbandgap state is induced as for the Carbon interstitial doping case. 
N doping should therefore lead to an increase in photocatalytic efficiency which again has shown by several experiments \cite{LIU2011,BUZBY2006,WANI2013}.
Looking at the calculated phase diagram, there is also a wide range of partial oxygen pressures where substitutional N doping shows to be energetically preferable, in contrast to Carbon. The magnetic coupling of two adjacent N atoms show to be AFM and like in the Carbon case and for to N atoms with greater distance to be paramagnetic. 
Paramagnetic behaviour for the N doping case, however, has already been observed in experiment \cite{LIU2011} and can be reproduced by our calculations. Recent experimental groups \cite{LIU2011,QIN2014,Drera2010} which investigated N-doped rutile single crystals, indeed measured ferromagnetic behaviour but cannot exclude the role of vacancies , created by the N ion implantation technique as possible sources of ferromagnetism.

\section*{Acknowledgements}
The authors acknowledge support from the Austrian Science Fonds FWF within  
SFB ViCoM F4109-N13 P09 (C.G. and P.M.). J.A. acknowledges the support of the ERC Advanced Grant 291414 "OxideSurfaces". We are indebted to Ulrike Diebold for most valuable and fruitful discussions.

\section*{References}
\end{document}